\documentclass[aps,twocolumn,pra,showpacs,amsmath,amssymb,showkeys,longbibliography,10pt]{revtex4-1}
\usepackage{graphicx}
\usepackage{epstopdf}
\usepackage{braket}
\usepackage{hyperref}
\allowdisplaybreaks

\begin{document}

\title{Coherence of resonant light-matter interaction in the strong-coupling limit}

\author{Th. K. Mavrogordatos}
\email[Email address(es): ]{themis.mavrogordatos@fysik.su.se;  th.mavrogordatos@gmail.com}
\affiliation{Department of Physics, Stockholm University, SE-106 91, Stockholm, Sweden}

\date{\today}

\begin{abstract}
We explore the role of quantum fluctuations in the strong-coupling limit of the dissipative Jaynes-Cummings oscillator driven on resonance. For weak excitation, we derive analytical expressions for the spectrum and the intensity correlation function for the photons scattered by the two-state atom coupled to the coherently driven cavity mode. We do so by writing down a birth-death process adding the higher orders in the excitation strength needed to go beyond the pure-state factorization, following the method introduced in [H. J. Carmichael, {\it Statistical Methods in Quantum Optics 2}, Springer, 2008, Sec. 16.3.4]. Our results for the first and second-order correlation functions are complemented by the numerical investigation of the waiting-time distribution for the photon emissions directed sideways, and the comparison with ordinary resonance fluorescence. To close out our discussion, we increase the driving field amplitude and approach the critical point organizing a second-order dissipative quantum phase transition by depicting the excitation pathways in the intracavity field distribution for a finite system size. 
\end{abstract}


\maketitle

\section{Introduction}

Single-atom quantum electrodynamics (QED) in its strong-coupling limit occupies a central position in the study of multi-photon quantum-nonlinear optics, providing the ground where the invalidation of the small-noise assumption, upon which the linear theory of fluctuations rests, is played out. Absorptive optical bistability for a single atom inside a resonant cavity is an elementary and illustrative example of a quantum dissipative system operating at the interface between the quantum and classical limit \cite{Savage1988}. The exquisite control acquired over cavity and circuit QED architectures has prompted an extensive investigation of the intrinsically quantum nonlinearity of coherent radiation-matter interaction most commonly encapsulated in the Jaynes-Cummings (JC) model \cite{JCpaper1963} and its extended versions (see e.g., \cite{Bishop2009}). In particular, the strong-coupling conditions attained in the circuit QED experiment of \cite{Fink2008} allowed the observation of vacuum Rabi splitting for up to two photons, demonstrating that `` the system is quantum mechanical in nature'' by going one step beyond the mode spitting that is in principle explainable by the treatment of two linear coupled oscillators. More recently, photon correlation functions of second and third order were employed in pursuit of an experimental signature of two-photon blockade, via the demonstration of three-photon antibunching with simultaneous two-photon bunching \cite{Hamsen2017}. 

Our interest here is with the resonant excitation of the JC oscillator in its strong-coupling limit. On resonance, photon blockade breaks down by means of a second-order dissipative quantum phase transition \textemdash{a} continuous transition organized around a critical point in the space of drive amplitude and detuning \cite{Carmichael2015}. It was as well in the strong-coupling limit that one of the earliest applications of quantum trajectory theory was given [see Sec. 5 of \cite{Alsing1991}], presenting a novel framework to account for the translation of microscopic spontaneous emission events to the accumulated macroscopic cavity-field switching effect as a diffusion process. Central to the demonstration of spontaneous dressed-state polarization is the emergence of new semiclassical states organizing the asymptotic dynamics (also called {\it neoclassical} states) when the length of the Bloch vector is conserved [see Sec. 4 of \cite{Alsing1991} and Sec. 16.3.1 for an extension of the Maxwell-Bloch solutions to $N$ atoms]. The quantum dissipative dynamics responsible for spontaneous symmetry breaking is elucidated by the demonstration of quantum jumps induced by the extraction of homodyne photocurrent records \textemdash{a} partial yet continuous observation \textemdash{in} the experiment of \cite{Armen2009} following the theoretical grounding provided by \cite{Mabuchi1998}. 

Instead of unravelling the density operator in quantum trajectories, the work reported in this brief Communication deals with ensemble-averaged quantities obtained from the master equation and the quantum regression formula. Initially, we carry on with the formalism developed in Sec. 16.3.4. of \cite{QO2} to account for the squeezing-induced linewidth narrowing in the fluorescence spectrum, as has been done for the transmitted light. After constructing an effective model within the secular approximation in the basis of dressed JC eigenstates for weak excitation in Sec. \ref{sec:MEsec}, we derive analytical results for the first and second-order correlation functions of atomic emission in Sec. \ref{sec:cohwel}. Finally, to reveal the role of quantum fluctuations close to the critical point, in Sec. \ref{sec:phasebmd} we consider a driving-field amplitude of the same order of magnitude as the light-matter coupling strength. We focus on the development of phase bimodality when the JC oscillator is driven such as to maintain a constant empty-cavity excitation. For this drive, the Maxwell-Bloch equations place us on the upper branch of the bistability curve, predicting that empty-cavity amplitude as a steady-state output. 

\section{Master equation, dressed states and reduced model}
\label{sec:MEsec}

Our starting point is the familiar Lindblad master equation (ME) for single-atom cavity QED with coherent driving of the cavity mode on resonance:
\begin{equation}\label{eq:ME1}
\begin{aligned}
 \frac{d\rho}{dt}&\equiv\mathcal{L}\rho=-i[\omega_0(\sigma_{+}\sigma_{-} + a^{\dagger}a)+ig(a^{\dagger}\sigma_{-}-a\sigma_{+}),\rho]\\
 &-i[\overline{\mathcal{E}}_0 e^{-i\omega_0 t}a^{\dagger} + \overline{\mathcal{E}}_0^{*} e^{i\omega_0 t}a,\rho]\\
 &+\kappa (2 a \rho a^{\dagger} -a^{\dagger}a \rho - \rho a^{\dagger}a)\\
 &+\frac{\gamma}{2}(2\sigma_{-}\rho \sigma_{+} - \sigma_{+}\sigma_{-}\rho - \rho \sigma_{+}\sigma_{-}),
 \end{aligned}
\end{equation}
where $\rho$ is the system density operator, $a$ ($a^{\dagger}$) are the annihilation (creation) operators for the cavity photons, and $\sigma_{+}$ ($\sigma_{-}$) are the raising (lowering) operators for the two-state atom dipole-coupled to the cavity mode with strength $g$. In what follows, this coupling constant is appreciably larger than the dissipation rates, $2\kappa$, which is the photon loss rate from the cavity, and $\gamma$, the rate at which the atom is damped to modes other than the privileged cavity mode, resonantly driven with amplitude $\overline{\mathcal{E}}_0$. The condition $g \gg \kappa, \gamma/2$ defines the strong-coupling limit, while the inequality $2(|\overline{\mathcal{E}}_0|/g) \ll 1$ defines the weak-driving regime which will concern us for the largest part of this Communication. We write Eq. \eqref{eq:ME1} in matrix form using the truncated basis $\{\ket{2}_A\ket{n}_a, \ket{1}_A\ket{n}_a; n=0,1,\ldots,N\}$ (the two states of the atom are denoted by the subscript $A$ and the Fock-states for the cavity field with the subscript $a$) and then solve the resulting set of linear ordinary differential equations using an explicit Runge-Kutta method of eighth order, checking the invariance of obtained results with respect to $N$. The steady states obtained have also been checked against the unique eigenstate corresponding to the zero eigenvalue of the Liouvillian super-operator. We have used the computational toolbox for quantum optics detailed in \cite{Tan1999} alongside codes developed ad hoc in the programming languages {\it Matlab} and {\it Python}. 

The stationary states of the resonantly driven JC model in the interaction picture \cite{DynamicStarkEffect, QO2} are the ``ground'' state
\begin{equation}\label{eq:groundstate}
 \ket{\tilde{\psi}_G}=S(\eta) \ket{O_{12}(r)}_A \ket{0}_a,
\end{equation}
with {\it quasi}energy
\begin{equation}\label{eq:quasienergiesJC1}
 E_G=0,
\end{equation}
and the ``excited'' state doublets ($n=1,2, \ldots$)
\begin{equation}
  \ket{\tilde{\psi}_{n,U (L)}}=D[\alpha(E_{n,U(L)})]S(\eta)\frac{1}{\sqrt{2}}\ket{\tilde{U}(\tilde{L})}
\end{equation}
with $\ket{\tilde{U}(\tilde{L})}\equiv \ket{O_{21}(r)}_A \ket{n-1}_a \pm i \ket{O_{12}(r)}_A \ket{n}_a$ (we take $U \to +, \, L \to -$) and {\it quasi}energies 
\begin{equation}\label{eq:quasienergiesJC2}
 E_{n,U(L)}=\pm e^{-3r} \sqrt{n} \hbar g.
 \end{equation}
The atomic-state superpositions, with coefficients depending on the degree of squeezing, $r$, have the form 
\begin{equation}
\begin{aligned}
  \ket{O_{12}(r)}_A \equiv &\frac{1}{\sqrt{2}}\Bigg(\sqrt{1 + e^{-2r}} \ket{1}_A \\
  &+ i e^{i{\rm arg}(\overline{\mathcal{E}}_0)} \sqrt{1 - e^{-2r}} \ket{2}_A\Bigg),
\end{aligned}
\end{equation}
\begin{equation}
\begin{aligned}
  \ket{O_{21}(r)}_A \equiv &\frac{1}{\sqrt{2}}\Bigg(\sqrt{1 + e^{-2r}} \ket{2}_A\\
  &- i e^{-i{\rm arg}(\overline{\mathcal{E}}_0)} \sqrt{1 - e^{-2r}} \ket{1}_A\Bigg),
\end{aligned}
\end{equation}
where $D(\alpha) \equiv \exp(\alpha a^{\dagger}-\alpha^{*}a)$ is the displacement operator with an energy-dependent argument 
$$\alpha(E;r)=-e^{i{\rm arg}(\overline{\mathcal{E}}_0)}[E/(\hbar g)] e^{4r}\sqrt{1-e^{-4r}};$$ $r$ is defined by
\begin{equation}
 e^{-2r} \equiv \sqrt{1 - (2|\overline{\mathcal{E}}_0|/g)^2}
\end{equation}
and features in the argument of the squeeze operator $S(\eta) \equiv \exp [\frac{1}{2}(\eta^{*}a^2-\eta a^{\dagger 2})]$ as $\eta=-r e^{2i {\rm arg}(\overline{\mathcal{E}}_0)}$. The states placed in quotes are stationary in the interaction and not in the Schr\"{o}dinger picture; they correspond to {\it quasi}energies and not actual energies. 

Equipped with the general properties of the dressed JC eigenstates, we will delineate the procedure for obtaining some analytical results in the weak-driving regime. We first make the approximation
\begin{equation}
 r=-\frac{1}{2}\ln\left[\sqrt{1-\left(2\frac{|\overline{\mathcal{E}}_0|}{g}\right)^2}\,\right] \simeq \left(\frac{|\overline{\mathcal{E}}_0|}{g}\right)^2,
\end{equation}
which is consistent with $\eta \approx - (\overline{\mathcal{E}}_0/g)^2$ and 
\begin{equation}
\begin{aligned}
 \alpha[E_{n,U(L)};r]& \approx -{\rm sgn}[E_{n,U(L)}] e^{i{\rm arg}(\overline{\mathcal{E}}_0)} \sqrt{n} (1+r) \sqrt{4r}\\
 &\approx -{\rm sgn}[E_{n,U(L)}] \sqrt{n} (2 \overline{\mathcal{E}}_0 /g),
 \end{aligned}
\end{equation}
for $2(|\overline{\mathcal{E}}_0|/g) \ll 1$. Further, we resort to the secular approximation in the basis of dressed JC eigenstates, writing the matrix elements of the density operator in the interaction picture as $\overline{\rho}_{E E^{\prime}}$, where $E, E^{\prime}$ are the {\it quasi}energies defined in Eqs. \eqref{eq:quasienergiesJC1}, \eqref{eq:quasienergiesJC2}. We adopt the notation used in Secs. 16.3.3 and 16.3.4 of \cite{QO2}. When transforming the ME \eqref{eq:ME1} in the dressed-state basis [considering only the nontrivial time evolution due to cavity and atomic damping], for the diagonal matrix elements we write down
\begin{equation}
 \dot{\overline{\rho}}_{E E}=\sum_{\epsilon} \left(\gamma_{\epsilon, E} \overline{\rho}_{\epsilon \epsilon} - \gamma_{E, \epsilon}\overline{\rho}_{E E}\right),
\end{equation}
with
\begin{equation}\label{eq:gammaratesdef}
 \gamma_{\epsilon, E}=\gamma \left|\braket{E|\sigma_{-}|\epsilon}\right|^2 + 2\kappa \left|\braket{E|a|\epsilon}\right|^2,
\end{equation}
while the off-diagonal elements obey the rate equations
\begin{equation}\label{eq:nondiag}
\begin{aligned}
  \dot{\overline{\rho}}_{E E^{\prime}}&=-\left(\sum_{\epsilon} \frac{\gamma_{E, \epsilon} + \gamma_{E^{\prime}, \epsilon}}{2} - K^{(1)}_{E,E^{\prime}}\right) \overline{\rho}_{E E^{\prime}}\\
  &+ K^{(2)}_{E,E^{\prime}} \overline{\rho}_{-E^{\prime}-E},
  \end{aligned}
\end{equation}
with
\begin{subequations}\label{eq:kapparatesdef}
 \begin{align}
   &K^{(1)}_{E,E^{\prime}} \equiv \gamma \braket{E|\sigma_{-}|E} \braket{E^{\prime}|\sigma_{+}|E^{\prime}} + 2\kappa \braket{E|a|E} \braket{E^{\prime}|a^{\dagger}|E^{\prime}}, \\
   &K^{(2)}_{E,E^{\prime}} \equiv \gamma \braket{E|\sigma_{-}|-E^{\prime}} \braket{-E|\sigma_{+}|E^{\prime}} \notag \\
   &+ 2\kappa \braket{E|a|-E^{\prime}} \braket{-E|a^{\dagger}|E^{\prime}}.
 \end{align}
\end{subequations}

The ``ground'' state of the time-independent JC Hamiltonian [see Eq. \eqref{eq:groundstate} and the approximations made in Eqs. 16.184-16.186 of \cite{QO2}], dressed by the weak drive, is approximated as
\begin{equation}\label{eq:gs}
 \ket{\tilde{\psi}_G}=\ket{G} + i\frac{\overline{\mathcal{E}}_0}{g}\ket{2}_A \ket{0}_a + \frac{1}{\sqrt{2}}\left(\frac{\overline{\mathcal{E}}_0}{g}\right)^2\ket{1}_A \ket{2}_a, 
\end{equation}
with $\ket{G} \equiv \ket{1}_A \ket{0}_a$, while the first ``excited'' doublet acquires a contribution from the ground state,
\begin{subequations}\label{eq:es}
 \begin{align}
 &\ket{\tilde{\psi}_{1,U}}=\ket{1,U} + i \frac{1}{\sqrt{2}}\frac{\overline{\mathcal{E}}^{*}_0}{g}\ket{1}_A \ket{0}_a,\\
 & \ket{\tilde{\psi}_{1,L}}=\ket{1,L} + i \frac{1}{\sqrt{2}}\frac{\overline{\mathcal{E}}^{*}_0}{g}\ket{1}_A \ket{0}_a,
 \end{align}
\end{subequations}
where 
\begin{subequations}
 \begin{align}
  \ket{1,U} \equiv \frac{1}{\sqrt{2}}(\ket{2}_A \ket{0}_a +i \ket{1}_A\ket{1}_a), \\
  \ket{1,L} \equiv \frac{1}{\sqrt{2}}(\ket{2}_A \ket{0}_a -i \ket{1}_A\ket{1}_a),
 \end{align}
\end{subequations}
is the first excited doublet of the JC ladder. Eqs. \eqref{eq:gs} and \eqref{eq:es} follow from a perturbative expansion of the dressed JC eigenstates \textemdash{linear} combinations of displaced and squeezed Fock states \cite{DynamicStarkEffect} \textemdash{in} powers of the ratio between the external driving-field amplitude and the light-matter coupling constant, since all operations upon which the treatment is based rely on the degree of squeezing. This treatment is followed in Sec. 16.3.4. of \cite{QO2} for the determination of the optical spectrum in the strong-coupling limit of cavity QED building upon the initial sketch of \cite{cavityQEDBerman}. As we will find out, all relevant observables and correlation functions we will be dealing with and this method can account for, are expressed in powers of $r \simeq (|\overline{\mathcal{E}}_0|/g)^2$.  
\begin{figure}
\begin{center}
\includegraphics[width=0.4\textwidth]{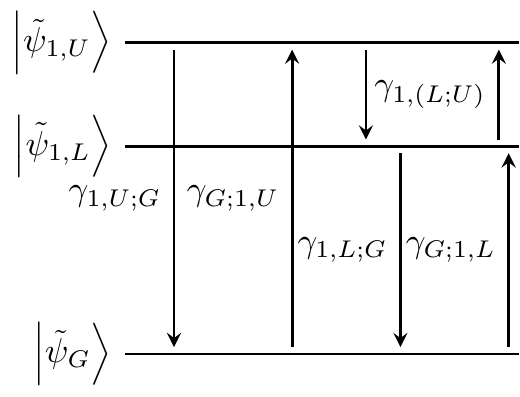}
\end{center}
\caption{{\it Schematic transition diagram in the effective three-level model.} Dominant scattering between dressed JC states at weak excitation for the diagonal elements of the reduced system density matrix [Eqs. \eqref{eq:gammarates}]; the transition rates out of the ``ground'' state, $\gamma_{G;1,U}, \gamma_{G;1,L}$ are a factor of $(|\overline{\mathcal{E}}_0|/g)^4$ smaller than those out of the ``excited'' states, $\gamma_{1,U;G}$ and $\gamma_{1,L;G}$. The transition rates within the ``excited'' state doublet are equal to each other.}
\label{fig:levels}
\end{figure}

Within the lowest-order approximation at weak excitation, we limit our attention to the transitions occurring between the ``ground'' state and the first ``excited'' state doublet, as depicted in Fig. \ref{fig:levels}. To specify the equations of motion for the matrix elements of the density matrix on the basis of these three dressed states, we use Eqs. \eqref{eq:gammaratesdef} and \eqref{eq:kapparatesdef} to determine the rates
\begin{subequations}\label{eq:gammarates}
 \begin{align}
&\gamma_{G;G}=\gamma|\braket{\tilde{\psi}_G|\sigma_{-}|\tilde{\psi}_G}|^2 + 2\kappa|\braket{\tilde{\psi}_G|a|\tilde{\psi}_G}|^2= \gamma \left(\frac{|\overline{\mathcal{E}}_0|}{g}\right)^2, \\
&\gamma_{G;1,U}=\gamma_{G;1,L}=(\kappa+\gamma/2)\left(\frac{|\overline{\mathcal{E}}_0|}{g}\right)^4, \\
&\gamma_{1,U;G}=\gamma_{1,L;G}=(\kappa + \gamma/2),
 \end{align}
\end{subequations}
together with
\begin{subequations}\label{eq:kapparates}
 \begin{align}
&K^{(1)}_{G;1,U}=K^{(1)}_{G;1,L}=K^{(1)}_{1,U;G}=K^{(1)}_{1,L;G}=\gamma\left(i\frac{\overline{\mathcal{E}}_0}{g}\right) \left(i\frac{1}{2}\frac{\overline{\mathcal{E}}^{*}_0}{g}\right) \notag \\
&=-\frac{\gamma}{2}\left(\frac{|\overline{\mathcal{E}}_0|}{g}\right)^2,\\
&K^{(2)}_{G;1,U}=K^{(2)}_{G;1,L}=\gamma\braket{\tilde{\psi}_G|\sigma_{-}|\tilde{\psi}_{1,L}} \braket{\tilde{\psi}_G|\sigma_{+}|\tilde{\psi}_{1,U}}\notag \\
&+2\kappa\braket{\tilde{\psi}_G|a|\tilde{\psi}_{1,L}}\braket{\tilde{\psi}_G|a^{\dagger}|\tilde{\psi}_{1,U}}=(\kappa + \gamma/2)\left(\frac{\overline{\mathcal{E}}^{*}_0}{g}\right)^2,\\
&K^{(2)}_{1,U;G}=K^{(2)}_{1,L;G}=(\kappa + \gamma/2)\left(\frac{\overline{\mathcal{E}}_0}{g}\right)^2.
 \end{align}
\end{subequations}
Hence, within the secular approximation, the rate equations for the diagonal elements are
\begin{subequations}
 \begin{align}
&\dot{\overline{\rho}}_{1,U;1,U}=-(\kappa + \gamma/2)\overline{\rho}_{1,U;1,U} + (\kappa + \gamma/2)\left(\frac{|\overline{\mathcal{E}}_0|}{g}\right)^4 \overline{\rho}_{GG},\\
&\dot{\overline{\rho}}_{1,L;1,L}=-(\kappa + \gamma/2)\overline{\rho}_{1,L;1,L} + (\kappa + \gamma/2)\left(\frac{|\overline{\mathcal{E}}_0|}{g}\right)^4 \overline{\rho}_{GG}, 
 \end{align}
\end{subequations}
with the stationary solution $\overline{\rho}_{1,U;1,U}=\overline{\rho}_{1,L;1,L}=(|\overline{\mathcal{E}}_0|/g)^4 \overline{\rho}_{GG}$. These matrix elements determine the steady-state density matrix (assuming $\overline{\rho}_{GG}=1$)
\begin{equation}\label{eq:rhoss}
  \overline{\rho}_{\rm ss}= |\tilde{\psi}_{G}\rangle \langle \tilde{\psi}_{G}| + \left(\frac{|\overline{\mathcal{E}}_0|}{g}\right)^4\left(|\tilde{\psi}_{1,U}\rangle \langle \tilde{\psi}_{1,U}| + |\tilde{\psi}_{1,L}\rangle \langle \tilde{\psi}_{1,L}| \right).
\end{equation}
going beyond the pure-state factorization ($\overline{\rho}_{\rm ss}= |\tilde{\psi}_{G}\rangle \langle \tilde{\psi}_{G}|$) by including a birth-death process contributing by higher-order terms to a mixed state. We also note that the three dressed states under consideration are normalized to dominant order, with $\braket{\tilde{\psi}_{G}|\tilde{\psi}_{1,(U,L)}}=0$ while $\braket{\tilde{\psi}_{1,(U,L)}|\tilde{\psi}_{1,(U,L)}}=(1/2)(|\overline{\mathcal{E}}_0|/g)^2$; the latter, however, combined with the ``excited'' state occupation probability $p_{1,\, {\rm ss}}=(|\overline{\mathcal{E}}_0|/g)^4$, leads to negligible contributions to the dominant terms in the steady-state observables extracted from the three-level model depicted in Fig. \ref{fig:levels}. Likewise, terms correcting the {\it ansatz} $\overline{\rho}_{GG}=1$, required by the normalization of $\overline{\rho}_{\rm ss}$, have a  negligible contribution.

As a first example based on the form of Eq. \eqref{eq:rhoss}, we note that the steady-state photon number $\braket{a^{\dagger}a}_{\rm ss}$ comprises two equal parts: one originates from the ``ground'' state (the last term on the right-hand side of Eq. \eqref{eq:gs}) and the other from the ``excited'' doublet, which is equal to $p_{1,\, {\rm ss}}$. Next, we briefly point to a characteristic property of atomic emission deriving from the form of \eqref{eq:rhoss}. We define $\braket{\tilde{\sigma}_{\pm}} \equiv {\rm tr}(\tilde{\rho}_{\rm ss} \sigma_{\pm})$, and compute the steady-state normal-ordered variance of the fluctuation $\Delta \tilde{\sigma}_{\theta} \equiv (1/2)(e^{-i\theta}\Delta\tilde{\sigma}_{-} + e^{i\theta}\Delta\tilde{\sigma}_{+})$ for an adjustable phase $\theta$ of the local oscillator employed in a common scheme for detecting squeezing \cite{Mandel1982}. We then find
\begin{equation}
\begin{aligned}
 \braket{:\left(\Delta \tilde{\sigma}_{\theta} \right)^2:}&=\frac{1}{2} \left(\frac{|\overline{\mathcal{E}}_0|}{g}\right)^2 \cos[2(\theta - {\rm arg}(\overline{\mathcal{E}}_0))] \\
 &+ \mathcal{O}[(|\overline{\mathcal{E}}_0|/g)^4].
 \end{aligned}
\end{equation}
Hence, squeezing of steady-state fluctuations attains its most negative value, $\braket{:(\Delta \tilde{\sigma}_{\theta=\pi/2 + {\rm arg}(\overline{\mathcal{E}}_0)})^2:} \simeq -r/2$, when the local oscillator is in phase with the mean polarization amplitude $\braket{\tilde{\sigma}_{-}}_{\rm ss} \approx i \overline{\mathcal{E}}_0/g$. The dependence of squeezing on the square of the driving field amplitude is also encountered in ordinary resonance fluorescence for weak excitation [where $|\overline{\mathcal{E}}_0|$ is instead scaled by $\gamma$, see Sec. 2.3.6 of \cite{QO1}]. 

\section{Coherence at weak excitation}
\label{sec:cohwel}

Having now extracted some indicative steady-state results in the weak excitation regime, let us see what can be learned about the coherence of the atomic emission from the effective three-level transition model, depicted in Fig. \ref{fig:levels}, originating from the master equation in the secular approximation.
\begin{figure*}
\begin{center}
\includegraphics[width=\textwidth]{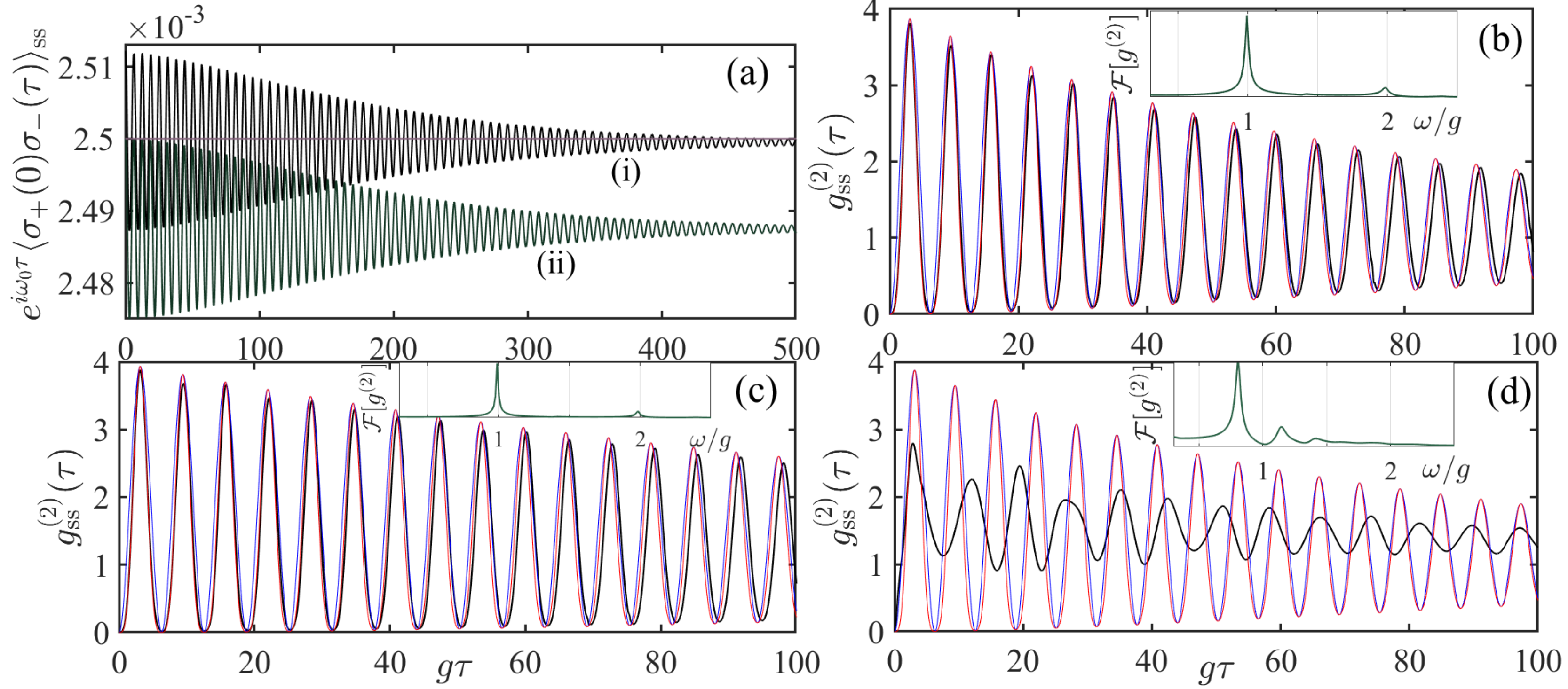}
\end{center}
\caption{{\it First and second-order coherence of fluorescence in the weak-excitation limit.} {\bf (a)} The numerically-evaluated first-order correlation function $e^{i\omega_0 \tau}\braket{\sigma_{+}(0)\sigma_{-}(\tau)}_{\rm ss}$ [curve (i)] is plotted against the analytical expression of Eq. \eqref{eq:firsorderCorr} [curve (ii) \textemdash{displaced} downwards by $(|\overline{\mathcal{E}}_0|/g)^4$ for visual clarity], for $\gamma/(2\kappa)=1$, $|\overline{\mathcal{E}}_0|/g=0.05$ and $g/\kappa=100$. The horizontal line marks the coherent component, $\braket{\tilde{\sigma}_{+}}_{\rm ss}\braket{\tilde{\sigma}_{-}}_{\rm ss}=(|\overline{\mathcal{E}}_0|/g)^2=0.0025$. Next, in frames {\bf (b)-(d)}, the second-order correlation function $g^{(2)}_{\rm ss}(\tau)$, extracted from the ME \eqref{eq:ME1} and the quantum regression formula [solid black curves], is plotted against the analytical expressions obtained from the JC three-level dressed-state formalism [Eq. \eqref{eq:g2DS}--thin blue lines] and the pure-state factorization method [Eq. \eqref{eq:g2PSfact}--thin red lines] for $g/\kappa=100$ and: $\gamma/(2\kappa)=1$, $|\overline{\mathcal{E}}_0|/g=0.05$ in {\bf (b)}, $\gamma/(2\kappa) \to 0$, $|\overline{\mathcal{E}}_0|/g=0.05$ in {\bf (c)}, $\gamma/(2\kappa)=1$, $|\overline{\mathcal{E}}_0|/g=0.25$ in {\bf (d)}. The three insets depict the modulus of the corresponding Fourier-transformed steady-state intensity correlation function. For all numerical results presented here we have taken $N=30$.}
\label{fig:g1g2plots}
\end{figure*}

\subsection{The fluorescence spectrum}

For this purpose, we turn to the off-diagonal matrix elements leading to dominant-order terms in $|\overline{\mathcal{E}}_0|/g$ (for instance, we do not take into account transitions within the ``excited'' state doublet). Invoking the quantum regression formula, the first-order correlation function required for the determination of the fluorescence spectrum is ($\tau \geq 0$) 
\begin{equation}\label{eq:g1general}
\begin{aligned}
 \braket{\sigma_{+}(0)\sigma_{-}(\tau)}_{\rm ss}&=e^{-i\omega_0 \tau} {\rm tr}[\sigma_{-} e^{\mathcal{\tilde{L}}\tau}(\tilde{\rho}_{\rm ss}\sigma_{+})]\\
 &=e^{-i\omega_0 \tau} {\rm tr}[\sigma_{-} \tilde{R}(\tau)],
 \end{aligned}
\end{equation}
with
\begin{equation}
 \tilde{R}(\tau) \equiv e^{\mathcal{\tilde{L}}\tau}(\tilde{\rho}_{\rm ss}\sigma_{+}),
\end{equation}
where $\mathcal{\tilde{L}}$ is the Liouvillian super-operator in the interaction picture. We therefore need to evaluate the matrix elements of $\tilde{R}(\tau)$ for all relevant $E, E^{\prime}$. To eliminate the trivial time dependence, as we did before, we transform
\begin{equation}
 \overline{R}_{EE^{\prime}} \equiv e^{(i/\hbar)(E-E^{\prime})t} \tilde{R}_{E E^{\prime}}.
\end{equation}
The initial values of the relevant matrix elements, which are to determine the dominant order in the perturbative expansion in powers of $(|\overline{\mathcal{E}}_0|/g)$, are
\begin{equation}
 \begin{cases}
  &(\tilde{\rho}_{\rm ss}\sigma_{+})_{G;1,U}=-\frac{1}{\sqrt{2}} \left(\displaystyle \frac{|\overline{\mathcal{E}}_0|}{g} \right)^2,  \\
  &(\tilde{\rho}_{\rm ss}\sigma_{+})_{1,U;G}=\frac{1}{\sqrt{2}} \left(\displaystyle \frac{|\overline{\mathcal{E}}_0|}{g}\right)^4, \\
    &(\tilde{\rho}_{\rm ss}\sigma_{+})_{G;1,L}=-\frac{1}{\sqrt{2}} \left( \displaystyle \frac{|\overline{\mathcal{E}}_0|}{g} \right)^2,\\
    &(\tilde{\rho}_{\rm ss}\sigma_{+})_{1,L;G}= \frac{1}{\sqrt{2}} \left(\displaystyle \frac{|\overline{\mathcal{E}}_0|}{g}\right)^4.
 \end{cases}
\end{equation}

At later times, $\tau >0$, the matrix elements written above satisfy the equations of motion corresponding to Eqs. \eqref{eq:nondiag} for the density-matrix elements:
\begin{subequations}\label{eq:generalSA}
 \begin{align}
 &\dot{\overline{R}}_{G;1,U}=-\left[\frac{1}{2}(\kappa + \gamma/2) - K^{(1)}_{G; 1,U}\right] \overline{R}_{G; 1,U} + K^{(2)}_{G;1,U} \overline{R}_{1,L;G}, \label{eq:generalSA1} \\
  &\dot{\overline{R}}_{G;1,L}=-\left[\frac{1}{2}(\kappa + \gamma/2) - K^{(1)}_{G; 1,L}\right] \overline{R}_{G; 1,L} + K^{(2)}_{G;1,L} \overline{R}_{1,U;G},\label{eq:generalSA2} \\
   &\dot{\overline{R}}_{1,U;G}=-\left[\frac{1}{2}(\kappa + \gamma/2) - K^{(1)}_{1,U;G}\right] \overline{R}_{1,U;G} + K^{(2)}_{1,U;G} \overline{R}_{G;1,L}, \label{eq:generalSA3} \\
    &\dot{\overline{R}}_{1,L;G}=-\left[\frac{1}{2}(\kappa + \gamma/2) - K^{(1)}_{1,L;G}\right] \overline{R}_{1,L;G} + K^{(2)}_{1,L;G} \overline{R}_{G;1,U} \label{eq:generalSA4} 
 \end{align}
\end{subequations}
and
\begin{subequations}
 \begin{align}
&\dot{\overline{R}}_{1,U(U);1,U(L)}=-(\kappa + \gamma/2)\overline{R}_{1,U(U);1,U(L)} \notag\\
&+ (\kappa + \gamma/2)\left(\frac{|\overline{\mathcal{E}}_0|}{g}\right)^4 \overline{R}_{G;G}, \\
&\dot{\overline{R}}_{1,L(L);1,L(U)}=-(\kappa + \gamma/2)\overline{R}_{1,L(L);1,L(U)} \notag \\
&+ (\kappa + \gamma/2)\left(\frac{|\overline{\mathcal{E}}_0|}{g}\right)^4 \overline{R}_{G;G},
 \end{align}
\end{subequations}

We note that in the set of Eqs. \eqref{eq:generalSA}, all coefficients $K^{(1)}_{E, E^{\prime}}$ are of order $(|\overline{\mathcal{E}}_0|/g)^2$ for all relevant $E, E^{\prime}$ values labeling the states involved in the dynamics within the interaction picture. Hence, we neglect their contribution to the damping rate $(1/2)(\kappa + \gamma/2)$. At the same time, the coefficients $K^{(2)}_{G;1,(U,L)}$ in Eqs. \eqref{eq:generalSA1} and \eqref{eq:generalSA2} multiply terms which are already of order higher than those on the left-hand side. Keeping then terms of the same order on both sides of the equations leads to
\begin{subequations}\label{eq:g1eqscoupled}
  \begin{align}
& \dot{\overline{R}}_{G;1,U}=-\frac{1}{2}(\kappa + \gamma/2)\overline{R}_{G;1,U} \\
& \dot{\overline{R}}_{G;1,L}=-\frac{1}{2}(\kappa + \gamma/2)\overline{R}_{G;1,L} \\
&\dot{\overline{R}}_{1,U;G}=-\frac{1}{2}(\kappa + \gamma/2)\overline{R}_{1,U;G} + (\kappa + \gamma/2) \left(\frac{|\overline{\mathcal{E}}_0|}{g}\right)^2 \overline{R}_{G;1,L},\\
    &\dot{\overline{R}}_{1,L;G}=-\frac{1}{2}(\kappa + \gamma/2) \overline{R}_{1,L;G} + (\kappa + \gamma/2) \left(\frac{|\overline{\mathcal{E}}_0|}{g}\right)^2 \overline{R}_{G;1,U}.
\end{align}
\end{subequations}
Their solutions read
\begin{subequations}\label{eq:Reqsfinal}
 \begin{align}
 &\overline{R}_{G;1,U}(\tau)=-\frac{1}{\sqrt{2}} \left(\displaystyle \frac{|\overline{\mathcal{E}}_0|}{g} \right)^2 e^{-\frac{1}{2}(\kappa + \gamma/2)\tau}, \\
  &\overline{R}_{G;1,L}(\tau)=-\frac{1}{\sqrt{2}} \left(\displaystyle \frac{|\overline{\mathcal{E}}_0|}{g} \right)^2 e^{-\frac{1}{2}(\kappa + \gamma/2)\tau}, \\
   &\overline{R}_{1,U;G}(\tau)=\frac{1}{\sqrt{2}}\left(\frac{|\overline{\mathcal{E}}_0|}{g}\right)^4 \{-1+2 [1 + (1/2)(\kappa + \gamma/2)\tau]\} \notag \\
   &\times e^{-\frac{1}{2}(\kappa + \gamma/2)\tau}, \\
   &\overline{R}_{1,L;G}(\tau)=\frac{1}{\sqrt{2}}\left(\frac{|\overline{\mathcal{E}}_0|}{g}\right)^4 \{-1+2 [1 + (1/2)(\kappa + \gamma/2)\tau]\} \notag \\
   & e^{-\frac{1}{2}(\kappa + \gamma/2)\tau}.  
\end{align}
\end{subequations}
It remains to find the matrix elements of $\sigma_{-}$ for the states involved in the transitions specified above:
\begin{equation}\label{eq:sigmame}
 \begin{cases}
  &\braket{\tilde{\psi}_G|\sigma_{-}|\tilde{\psi}_{1,U}}=1/\sqrt{2},\\
  &\braket{\tilde{\psi}_{1,U}|\sigma_{-}|\tilde{\psi}_{G}}=-(1/\sqrt{2})(|\overline{\mathcal{E}}_0|/g)^2, \\
  &\braket{\tilde{\psi}_G|\sigma_{-}|\tilde{\psi}_{1,L}}=1/\sqrt{2}, \\
  &\braket{\tilde{\psi}_{1,L}|\sigma_{-}|\tilde{\psi}_{G}} =-(1/\sqrt{2})(|\overline{\mathcal{E}}_0|/g)^2. 
 \end{cases}
\end{equation}
Substituting the expressions of Eqs. \eqref{eq:Reqsfinal} and \eqref{eq:sigmame} into Eq. \eqref{eq:g1general}, we arrive at
\begin{equation}
\begin{aligned}
 e^{i\omega_0 \tau} \braket{\sigma_{+}(0)\sigma_{-}(\tau)}_{\rm ss}=&\braket{\tilde{\psi}_G|\sigma_{-}|\tilde{\psi}_{G}} \overline{R}_{G;G} \notag \\
 &+\braket{\tilde{\psi}_G|\sigma_{-}|\tilde{\psi}_{1,U}} \overline{R}_{1,U;G}(\tau) e^{-ig\tau} \notag \\
 &+\braket{\tilde{\psi}_G|\sigma_{-}|\tilde{\psi}_{1,L}} \overline{R}_{1,U;L}(\tau)e^{ig\tau}\notag \\
 &+ \braket{\tilde{\psi}_{1,U}|\sigma_{-}|\tilde{\psi}_{G}} \overline{R}_{G;1,U}(\tau)e^{ig\tau} \notag \\
 &+  \braket{\tilde{\psi}_{1,L}|\sigma_{-}|\tilde{\psi}_{G}} \overline{R}_{G;1,L}(\tau)e^{-ig\tau}
 \end{aligned}
\end{equation}
and, finally, it follows that ($\tau \geq 0$)
\begin{equation}\label{eq:firsorderCorr}
\begin{aligned}
 &\braket{\sigma_{+}(0)\sigma_{-}(\tau)}_{\rm ss}=\left(\frac{|\overline{\mathcal{E}}_0|}{g}\right)^2 e^{-i\omega_0 \tau} +  2 \left(\frac{|\overline{\mathcal{E}}_0|}{g}\right)^4\\
 & \times e^{-[\frac{1}{2}(\kappa + \gamma/2)+i\omega_0]\tau} \left[1 + \frac{1}{2}(\kappa + \gamma/2)\tau\right] \cos(g\tau).
 \end{aligned}
\end{equation}

The numerically-evaluated first-order correlation function $\braket{\sigma_{+}(0)\sigma_{-}(\tau)}_{\rm ss}$ is depicted in Fig. \ref{fig:g1g2plots}(a) and compared to the analytical expression on Eq. \eqref{eq:firsorderCorr}. The level at $(|\overline{\mathcal{E}}_0|/g)^2$, which is the coherent part of the spectrum, coincides with the prediction of the Maxwell-Bloch equations for the product $\braket{\sigma_{+}}_{\rm ss} \braket{\sigma_{-}}_{\rm ss}$ in the lower branch of the absorptive bistability curve when $\gamma \to 0$, as well as with the corresponding neoclassical expressions \cite{Alsing1991, QO2}. The deviation caused by quantum fluctuations coincides with the first-order correlation function of the intracavity field [Eq. (16.203) of \cite{QO2}]. This can be anticipated from the similar form of the expressions for $a^{\dagger}a=(1/2)(|1,U \rangle \langle 1,U| +|1,L \rangle \langle 1,L| +|1,U \rangle \langle 1,L|+ |1,L \rangle \langle 1,U|)$ and $\sigma_{+}\sigma_{-}=(1/2)(|1,U \rangle \langle 1,U| +|1,L \rangle \langle 1,L| -|1,U \rangle \langle 1,L|- |1,L \rangle \langle 1,U|)$ within the effective three-level model (in its ``bare'' JC-dressed form) we are here considering \cite{Shamailov2010}. 

The fluorescence spectrum, then, to dominant order in $|\overline{\mathcal{E}}_0|/g$ and including both coherent and incoherent parts, reads
\begin{equation}
 \begin{aligned}
T(\omega)&=\frac{1}{2\pi}\int_{-\infty}^{\infty}d\tau \exp(i\omega\tau)\braket{\sigma_{+}(0)\sigma_{-}(\tau)}_{\rm ss}\\
&=\frac{1}{\pi}{\rm Re}\left[\int_{0}^{\infty}d\tau \exp(i\omega\tau)\braket{\sigma_{+}(0)\sigma_{-}(\tau)}_{\rm ss}\right]\\
&=\left(\frac{|\overline{\mathcal{E}}_0|}{g}\right)^2\delta(\omega-\omega_0) + \left(\frac{|\overline{\mathcal{E}}_0|}{g}\right)^4 \frac{2}{\pi}\\
&\times \Bigg\{\frac{[\frac{1}{2}(\kappa + \gamma/2)]^3}{\{[\frac{1}{2}(\kappa + \gamma/2)]^2 + (\omega-\omega_0 + g)^2\}^2}\\
&+\frac{[\frac{1}{2}(\kappa + \gamma/2)]^3}{\{[\frac{1}{2}(\kappa + \gamma/2)]^2 + (\omega-\omega_0 - g)^2\}^2}\Bigg\}.
 \end{aligned}
\end{equation}
As we expect, the incoherent part of the spectrum coincides with the spectrum of the transmitted light, both revealing a vacuum Rabi doublet with squeezing-induced linewidth narrowing. The coefficient of the delta function in the coherent part of the spectrum coincides with the prediction of the Maxwell-Bloch equations for $\gamma \to 0$ and $|\overline{\mathcal{E}}_0|/g \leq 1/(2\sqrt{2})$, namely $\braket{\tilde{\sigma}_{+}}_{\rm ss}\braket{\tilde{\sigma}_{-}}_{\rm ss}=(|\overline{\mathcal{E}}_0|/g)^2$. For the transmitted light, the mean-field prediction in that limit is $\braket{\tilde{a}}_{\rm ss}=\braket{\tilde{a}^{\dagger}}_{\rm ss}=0$, consistent with the absence of a coherent part in the spectrum within our perturbative treatment in powers of $|\overline{\mathcal{E}}_0|/g$. 

\subsection{Intensity correlation function and waiting-time distribution}

We will now extend our analysis to the second-order coherence of fluorescence. The intensity correlation function for the photons scattered by the two-level atom is defined as
\begin{equation}\label{eq:g2A}
\begin{aligned}
 g^{(2)}_{\rm ss}(\tau)& \equiv \frac{\braket{\sigma_{+}(0)\sigma_{+}(\tau)\sigma_{-}(\tau)\sigma_{-}(0)}_{\rm ss}}{\braket{\sigma_{+}\sigma_{-}}_{\rm ss}^2}\\
 &=\frac{{\rm tr} \left\{[e^{\tilde{\mathcal{L}}\tau}\rho_{\rm cond}] \sigma_{+}\sigma_{-}\right\}}{\braket{\sigma_{+}\sigma_{-}}_{\rm ss}}\\
 &={\rm tr} [\sigma_{+}\sigma_{-} \tilde{D}(\tau)],
 \end{aligned}
\end{equation}
where $\rho_{\rm cond}\equiv (\sigma_{-}\rho_{\rm ss}\sigma_{+})/[{\rm tr}(\sigma_{-}\rho_{\rm ss}\sigma_{+})]$ is the conditional density matrix following the emission of one photon from the atom, evolving in time as $\tilde{D}(\tau) \equiv e^{\tilde{\mathcal{L}}\tau}\rho_{\rm cond}$. Likewise, we first compute the matrix elements of $\rho_{\rm cond}$ in the dressed-state basis. We find,
\begin{equation}
\begin{aligned}
 &\overline{D}_{1,(U,L);G}\equiv \braket{\tilde{\psi}_{1,(U,L)}|\sigma_{-}\rho_{\rm ss}\sigma_{+}|\tilde{\psi}_{G}}\\
 &=\langle\tilde{\psi}_{1,(U,L)}|\sigma_{-}\rho_{\rm ss}| 2 \rangle_{A} |0 \rangle_{a}\\
 &=-i\frac{1}{\sqrt{2}} \frac{\overline{\mathcal{E}}_0}{g} \left[ \left(\frac{|\overline{\mathcal{E}}_0|}{g}\right)^2 + \left(\frac{|\overline{\mathcal{E}}_0|}{g}\right)^4 \right].
 \end{aligned}
\end{equation}
Similarly,
\begin{equation}
\begin{aligned}
 &\overline{D}_{G; 1,(U,L)}\equiv \braket{\tilde{\psi}_{G}|\sigma_{-}\rho_{\rm ss}\sigma_{+}|\tilde{\psi}_{1,(U,L)}}\\
 &=\frac{i}{\sqrt{2}}\frac{\overline{\mathcal{E}}_0^{*}}{g}\langle\tilde{\psi}_{G}|\sigma_{-}\rho_{\rm ss}| 2 \rangle_{A} |0 \rangle_{a}\\
 &=\frac{i}{\sqrt{2}}\frac{\overline{\mathcal{E}}_0^{*}}{g}\left[ \left(\frac{|\overline{\mathcal{E}}_0|}{g}\right)^2 + \left(\frac{|\overline{\mathcal{E}}_0|}{g}\right)^4 \right].
 \end{aligned}
\end{equation}
As for the matrix elements of the conditional density matrix within the same family of dressed states, we have
\begin{equation}
  \overline{D}_{G;G} \equiv \braket{\tilde{\psi}_{G}|\sigma_{-}\rho_{\rm ss}\sigma_{+}|\tilde{\psi}_{G}}=\left(\frac{|\overline{\mathcal{E}}_0|}{g}\right)^2 + \left(\frac{|\overline{\mathcal{E}}_0|}{g}\right)^4
\end{equation}
and
\begin{equation}
\begin{aligned}
 &\overline{D}_{1,(U,L); 1,(U,L)}\equiv\braket{\tilde{\psi}_{1,(U,L)}|\sigma_{-}\rho_{\rm ss}\sigma_{+}|\tilde{\psi}_{1,(U,L)}}\\
 &=\frac{i}{\sqrt{2}}\frac{\overline{\mathcal{E}}_0^{*}}{g}\langle\tilde{\psi}_{1,(U,L)}|\sigma_{-}\rho_{\rm ss}| 2 \rangle_{A} |0 \rangle_{a}\\
 &\approx \left(\frac{i}{\sqrt{2}}\frac{\overline{\mathcal{E}}_0^{*}}{g}\right)\langle\tilde{\psi}_{1,(U,L)}|\sigma_{-}|\tilde{\psi}_{G}\rangle \langle \tilde{\psi}_{G}| 2 \rangle_{A} |0 \rangle_{a} =\frac{1}{2}\left(\frac{|\overline{\mathcal{E}}_0|}{g}\right)^4.
 \end{aligned}
\end{equation}
The matrix elements for the two-level excitation operator read
\begin{equation}
 \begin{cases}
  &\braket{\tilde{\psi}_G|\sigma_{+}\sigma_{-}|\tilde{\psi}_{1,U}}=-(i/\sqrt{2})(\overline{\mathcal{E}}_0^{*}/g), \\ &\braket{\tilde{\psi}_{1,U}|\sigma_{+}\sigma_{-}|\tilde{\psi}_{G}}=(i/\sqrt{2})(\overline{\mathcal{E}}_0/g), \\
  &\braket{\tilde{\psi}_G|\sigma_{+}\sigma_{-}|\tilde{\psi}_{1,L}}=-(i/\sqrt{2})(\overline{\mathcal{E}}_0^{*}/g), \\ &\braket{\tilde{\psi}_{1,L}|\sigma_{+}\sigma_{-}|\tilde{\psi}_{G}} =(i/\sqrt{2})(\overline{\mathcal{E}}_0/g), \\
  &\braket{\tilde{\psi}_G|\sigma_{+}\sigma_{-}|\tilde{\psi}_{G}}=(|\overline{\mathcal{E}}_0|/g)^2, \\ &\braket{\tilde{\psi}_{1,(U,L)}|\sigma_{+}\sigma_{-}|\tilde{\psi}_{1,(U,L)}}=1/2.
 \end{cases}
\end{equation}

Now, the matrix elements accounting for the excitation and de-excitation of the first ``excited'' states obey one and the same equation of motion, since the second terms in the sums of the right-hand side for the four equations corresponding to the set \eqref{eq:generalSA} (with $R \to D$) only contribute further higher-order corrections. In particular, the equations in question reduce to
\begin{equation}\label{eq:melbetg2}
 \dot{\overline{D}}_{G (1,[U,L]); 1,[U,L] (G)}=-\frac{1}{2}(\kappa + \gamma/2)\overline{D}_{G (1,[U,L]); 1,[U,L] (G)},
\end{equation}
giving rise to a simple exponential decay of correlations. Putting all these pieces together, we obtain
\begin{equation}\label{eq:g2DS}
 g^{(2)}_{\rm ss}(\tau)=1+e^{-(\kappa+\gamma/2)\tau}-2e^{-\frac{1}{2}(\kappa+\gamma/2)\tau}\cos(g\tau).
\end{equation}
For a single atom in a cavity, the pure-state factorization within the two-quanta basis for one atom in a cavity yields the following correlation function of the side-scattered light in the weak excitation limit:
\begin{equation}\label{eq:g2PSfact}
\begin{aligned}
 & g^{(2)}_{\rm ss}(\tau)=\Bigg\{1-e^{-\frac{1}{2}(\kappa+\gamma/2)\tau} \\
  &
\times \left[\cos(g^{\prime}\tau) + \frac{\kappa-\gamma^{\prime}/2}{\kappa+\gamma/2}\frac{\frac{1}{2}(\kappa + \gamma/2)}{g^{\prime}} \sin(g^{\prime}\tau)\right]\Bigg\}^2.
\end{aligned}
\end{equation}
where
$g^{\prime}\equiv \sqrt{g^2-\frac{1}{4}(\kappa-\gamma/2)^2}$ and $\gamma^{\prime}\equiv \gamma (1+2C_1)$ is the cavity-enhanced emission rate featuring the (single-atom) co-operativity parameter $2C_1 \equiv 2g^2/(\kappa \gamma)$. In the strong-coupling limit, $g \gg \kappa, \gamma/2$, the two expressions of Eqs. \eqref{eq:g2DS} and \eqref{eq:g2PSfact} \textemdash{both} independent of the drive strength \textendash{practically} coincide as we can see in frames (b)-(d) of Fig. \ref{fig:g1g2plots}. The former expression is an expansion of the square in the latter when ignoring the second term in the square brackets. The closeness between the two expressions in the strong-coupling limit may be traced to the fact that the ``ground'' state of Eq. \eqref{eq:gs} agrees with the pure state in the two-quanta basis [see Eq. (2) of \cite{Carmichael1991}] and it is the coefficient of $\ket{2}_A \ket{0}_a$, of order $|\overline{\mathcal{E}}_0|/g$ that ultimately determines the intensity correlation function. The coefficients $\tilde{\alpha}(t)$ and $\tilde{\beta}(t)$ featuring in the one-quantum expansion $\ket{\tilde{\psi}(\tau)}=\ket{1}_{A}\ket{0}_a + \tilde{\alpha}(\tau) \ket{1}_{A}\ket{1}_a + \tilde{\beta}(\tau)\ket{2}_{A}\ket{0}_a,$ obey linear coupled oscillator equations solved by the familiar vacuum Rabi oscillations between the one-quantum amplitudes. Since there is a separation of orders in Eqs. \eqref{eq:melbetg2} determining the transitions between the ``ground'' and ``excited'' states, the damped Rabi oscillation is essentially all what remains; in that sense, there is nothing from Eq. \eqref{eq:g2DS} which is inaccessible to the theory of linear coupled oscillators.
\begin{figure}
\begin{center}
\includegraphics[width=0.46\textwidth]{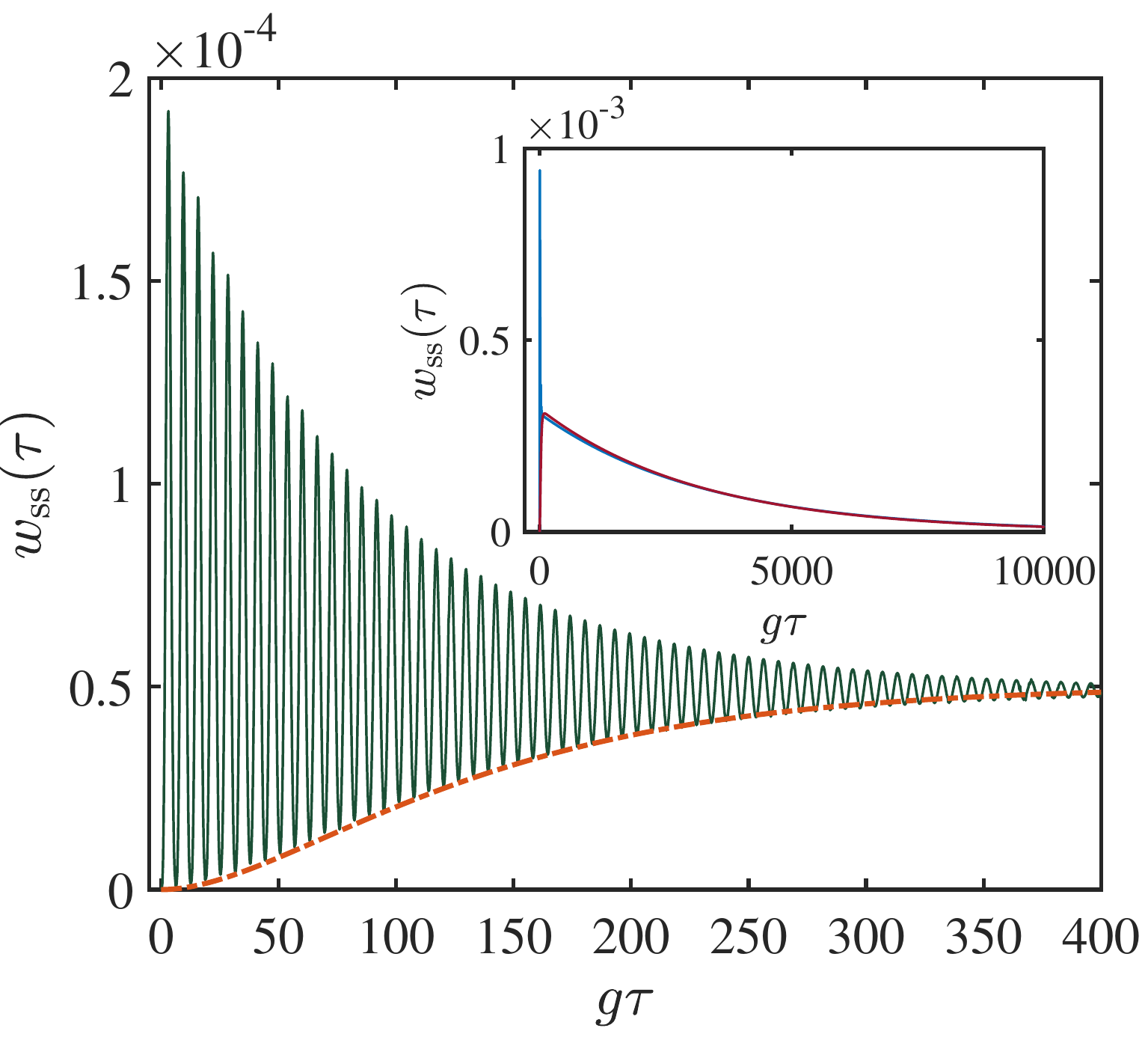}
\end{center}
\caption{{\it Detecting photons scattered from an atom driven by a cavity mode vs. free-space resonance fluorescence.} The initial part of the evolving waiting-time distribution $w_{\rm ss}(\tau)$ between side emissions from a two-level atom coupled to a resonant cavity mode with $\gamma/(2\kappa)=1$, $|\overline{\mathcal{E}}_0|/g=0.05$ and $g/\kappa=100$, is plotted [in solid blue] against the waiting-time distribution of ordinary resonance fluorescence with an effective drive amplitude $\overline{\mathcal{E}}_0^{\prime}=g \braket{a}_{\rm ss}$, from Eq. \eqref{eq:resflwt} [in dot-dashed orange]. The inset depicts the full evolution of the two waiting-time distributions [the blue curve depicts numerical results -- Eq. \eqref{eq:wss}, and the purple curve depicts the corresponding analytical expression obtained from the theory of resonance fluorescence -- Eq. \eqref{eq:resflwt}] when operating away from the strong-coupling limit, with $\gamma/(2\kappa)=1/2$, $|\overline{\mathcal{E}}_0|/g=0.05$ and $g/\kappa=8$. For all numerical results presented here we have taken $N=25$.}
\label{fig:wtau}
\end{figure}
\begin{figure*}
\begin{center}
\includegraphics[width=0.85\textwidth]{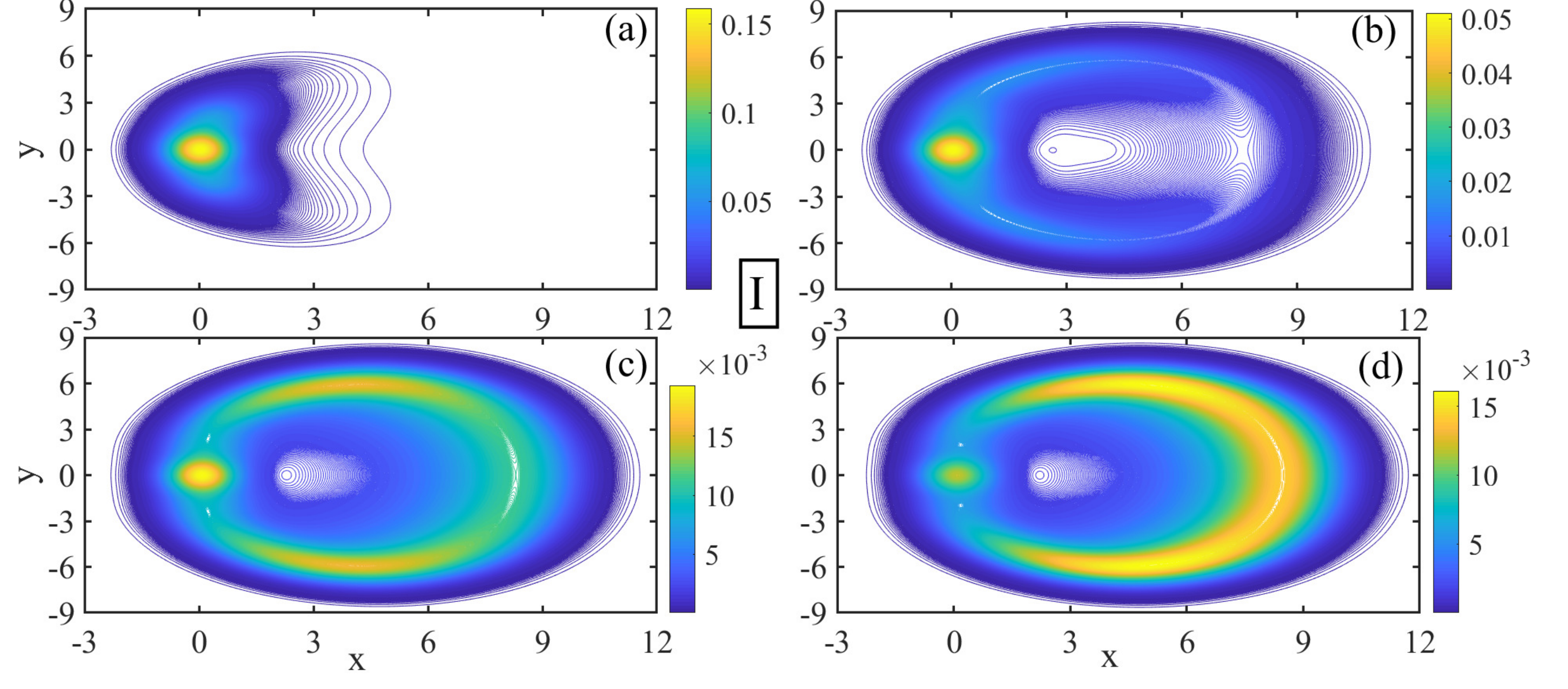}
\includegraphics[width=0.85\textwidth]{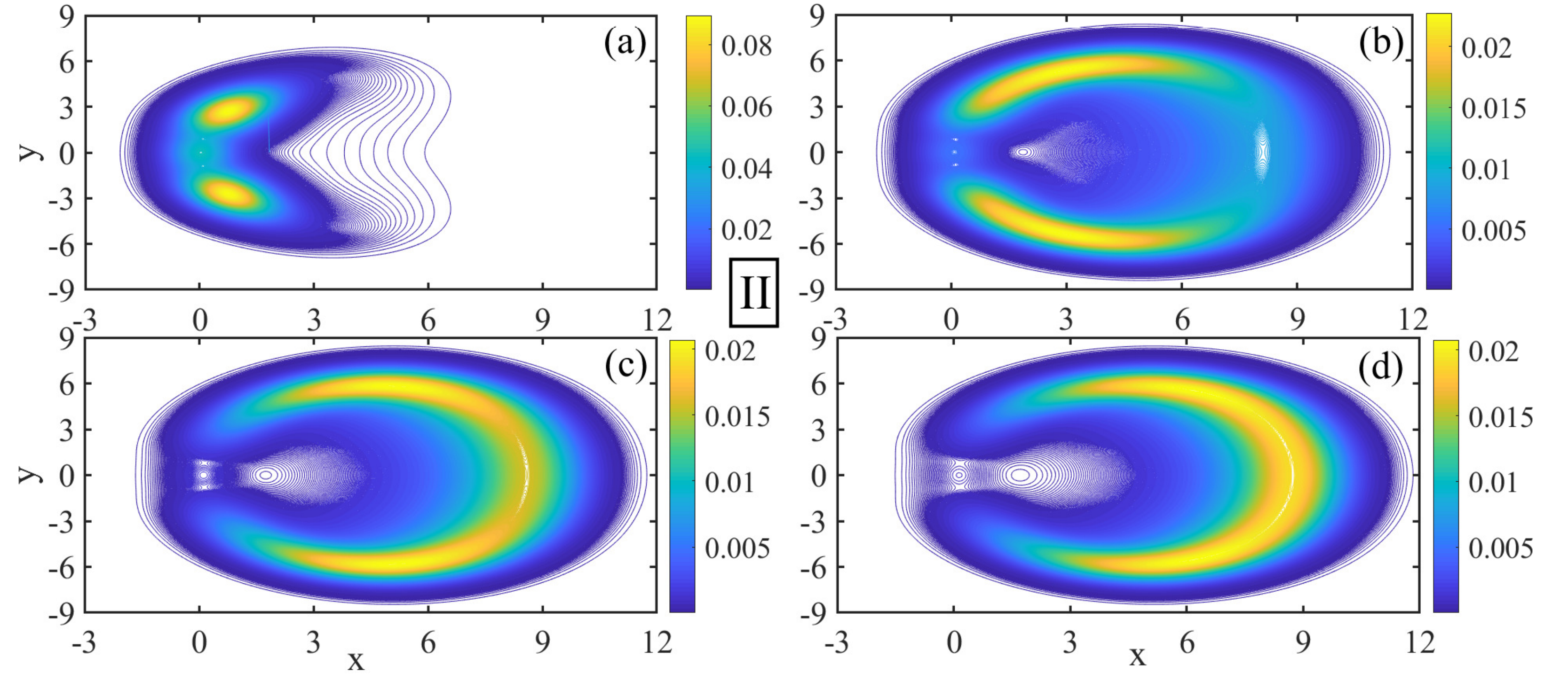}
\end{center}
\caption{{\it Approaching the critical point of a second-order dissipative quantum phase transition for a finite system size.} Contour plots of $Q_{\rm ss}(x+iy)$ for a fixed ratio $|\overline{\mathcal{E}}_0|/\kappa=10$ when driving the cavity mode below, $|\overline{\mathcal{E}}_0|/g=0.45$, in {\bf Panel I}, and at the critical point, $|\overline{\mathcal{E}}_0|/g=0.5$, in {\bf Panel II}. The ratio of the two dissipation rates, $\gamma/(2\kappa)$, is equal to: $0$ in {\bf (a)}, $1$ in {\bf (b)}, $5/3$ in {\bf (c)} and $2$ in {\bf (d)}, across both panels. For all numerical results presented here we have taken $\overline{\mathcal{E}}_0=i|\overline{\mathcal{E}}_0|$ and $N=200$.}
\label{fig:Qfuncs}
\end{figure*}
\begin{figure}
\begin{center}
\includegraphics[width=0.5\textwidth]{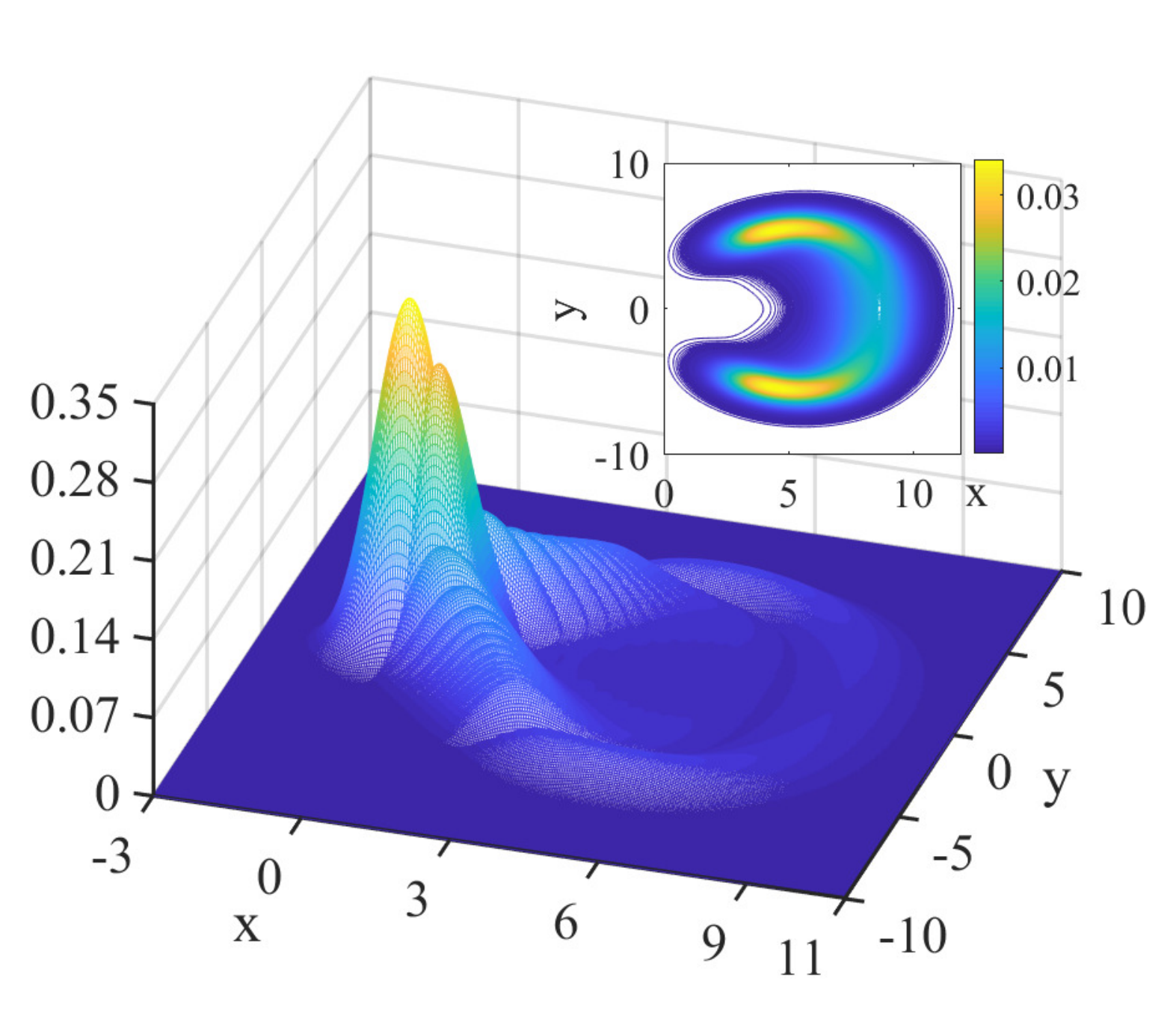}
\end{center}
\caption{{\it Transient intracavity field distribution above the critical point.} Surface plots of $Q[\tilde{\rho}_c(t_n)](x+iy)$, generated from the formula corresponding to Eq. \eqref{eq:Qdef} for a given instant of time, for twelve system density operators solving the ME \eqref{eq:ME1} at equidistant times $t_n$, $n=1,2,\ldots,12$, spanning the interval $gT=[0, 27.5]$ in the course of the evolution from the ground (product) state $\tilde{\rho}(0)=({}_a|0\rangle \langle 0|_a) \otimes ({}_A|1\rangle \langle 1|_A)$. The surface plots are superimposed on top of each other as well as on top of the steady-state {\it quasi}probability distribution. A contour plot of the latter is also given in the inset for reference. The operating conditions read: $|\overline{\mathcal{E}}_0|/g=0.6$, $\gamma/(2\kappa)=1$ and $g/\kappa=50/3$. For all numerical results presented here we have taken $\overline{\mathcal{E}}_0=i|\overline{\mathcal{E}}_0|$ and $N=120$.}
\label{fig:Qfuncevol}
\end{figure}

This reduction, however, provides a limited perspective: the correlation function of Eq. \eqref{eq:g2PSfact} extends beyond the strong-coupling limit unlike its counterpart, as long as we remain in the weak-excitation regime. When we operate closer the critical point, there is a obvious deviation between the numerically-evaluated intensity correlation function and the prediction of Eqs. \eqref{eq:g2DS} and \eqref{eq:g2PSfact}, first and foremost in terms of their frequency content; the main spectral peak in the Fourier transform of $g_{\rm ss}^{(2)}(\tau)$ splits on either side of $\omega=g$ [see Fig. \ref{fig:g1g2plots}(d)]. When we increase $|\overline{\mathcal{E}}_0|/g$ further, oscillations at $\omega \sim g$ fade away in favor of a more collective response involving competing states spanning the two distinct branches of the split JC ladder \cite{Chough1996}.

To conclude this section on second-order coherence, we note that the current scheme does not allow for the calculation of the intensity correlation function, since terms of order $(|\overline{\mathcal{E}}_0|/g)^8$ are involved \textemdash{this} observation brings in the inclusion of the second ``excited''-state doublet. Nevertheless, we find that a rough estimate of $g^{(2)}_{{\rm ss}, \rightarrow}(0)$ coming from $\ket{\tilde{\psi}_G}$,
\begin{equation}\label{eq:g2forward}
 \frac{\braket{\tilde{\psi}_G|(a^{\dagger})^2a^2|\tilde{\psi}_G}}{\braket{\tilde{\psi}_G|a^{\dagger}a|\tilde{\psi}_G}^2}=\frac{1}{4} \left(\frac{g}{|\overline{\mathcal{E}}_0|}\right)^4=\frac{1}{4r^2},
\end{equation}
is reasonably close to the result obtained numerically from the full solution of the ME, predicting the presence of significant photon bunching. For a large spontaneous emission enhancement factor $2C_1 \equiv 2g^2/(\kappa \gamma)$, the forwards photon scattering may also be highly bunched in the weak-excitation regime of the bad-cavity limit, with $g_{\rightarrow}^{(2)}(0)=(1-4C_1^2)^2$ \cite{RiceCarmichaelIEEE}, as well as for the forwards-scattering channel comprising the emission of a free two-state atom excited by the output of a coherently driven cavity with low flux in a cascaded open quantum systems formulation \cite{Carmichael1993}. In both these instances, however, the zero-delay intensity correlation function is independent of the drive amplitude in the limit $|\overline{\mathcal{E}}_0| \to 0$, unlike the asymptotic behavior anticipated by Eq. \eqref{eq:g2forward} for the strong-coupling limit.

Next, let us trace how is the coherence associated with a well defined JC ladder manifested in a closely related quantity to $g^{(2)}_{\rm ss}(\tau)$, the photoelectron waiting-time distribution; this is the distribution of time intervals $\tau$ between two successive photoelectric detection events. The waiting-time distribution for side-scattered photons at unit detection efficiency is defined as \cite{QO1} 
\begin{equation}\label{eq:wss}
 w_{\rm ss}(\tau) \equiv \gamma \frac{{\rm tr}[\sigma_{+}\sigma_{-}\, e^{\overline{\mathcal{L}}\tau}(\sigma_{-}\rho_{\rm ss} \sigma_{+})]}{\braket{\sigma_{+}\sigma_{-}}_{\rm ss}},
\end{equation}
where $\overline{\mathcal{L}}\equiv \mathcal{L}-\gamma \sigma_{-} \cdot \sigma_{+}$, imposing the condition that no photons are emitted sideways in the interval $\tau$. It is instructive to compare the numerically computed function from Eq. \eqref{eq:wss} against the waiting-time distribution of ordinary resonance fluorescence from a free atom driven by a coherent field. Such a field is to be assigned an effective amplitude $\overline{\mathcal{E}}_0^{\prime}=g \braket{a}_{\rm ss}$, where $\braket{a}_{\rm ss}$ is the steady-state intracavity amplitude numerically obtained from the ME \eqref{eq:ME1}:
\begin{equation}\label{eq:resflwt}
  \tilde{w}_{\rm ss}(\tau)=\gamma e^{-(\gamma/2)\tau}\frac{Y^2}{1-2Y^2}\left[1-\cosh\left(\frac{\gamma \tau}{2}\sqrt{1-2Y^2}\right)\right].
\end{equation}
In Eq. \eqref{eq:resflwt}, $Y \equiv 2\sqrt{2} |\overline{\mathcal{E}}_0^{\prime}|/\gamma$ is the dimensionless drive parameter the square of which determines the ratio of the incoherent to the coherent scattering intensities. For the weakly resonantly driven JC oscillator, on the other hand, this ratio is solely determined by $(|\overline{\mathcal{E}}_0|/g)^2$. In Fig. \ref{fig:wtau}, we observe that the decaying Rabi oscillations in $w_{\rm ss}(\tau)$ are bounded from below by $\tilde{w}_{\rm ss}(\tau)$ plotted for $Y \simeq 0.07$, with a peak occurring at about $\gamma \tau=12$. Increasing the dissipation rates by one order of magnitude, we find that the effective ordinary resonance fluorescence model maintains the same value of $Y^2 \ll 1$ and therefore the same position where $\tilde{w}_{\rm ss}(\tau)$ attains its maximum. However, Rabi oscillations survive only up to $g\tau \simeq 50$. As we move away from the strong-coupling limit at weak excitation, numerical simulations show that the mean time interval between photopulses is in good agreement with the prediction of the theory for resonance fluorescence, namely $\tau_{\rm av}=(2/\gamma)(1+Y^2)/Y^2 \simeq 2/(\gamma Y^2)$. Nevertheless, the initial high-frequency oscillations at $g$ produce a continual disagreement with respect to the position and the value of the maximum in the probability distribution \textemdash{a} distinct mark left by the dressed JC energy levels.

\section{Organization of phase bimodality for a finite system size}
\label{sec:phasebmd}

We will close our discussion on the coherence properties of resonant light-matter interaction by making a brief detour to the onset of the well-known dissipative quantum phase transition associated with the critical point $|\overline{\mathcal{E}}_0|=g/2$ and the emergence of spontaneous dressed state polarization in the limit of ``zero system size'' $\gamma \to 0$~\cite{Carmichael2015}. We will focus specifically on the distribution of the intracavity field for increasing values of the ratio $\gamma/(2\kappa)$ as we approach the critical point from below. To that end, in Fig. \ref{fig:Qfuncs} we plot the steady-state {\it quasi}probability distribution $Q_{\rm ss}(\alpha)$ (with $\alpha=x+iy$) calculated from \cite{Alsing1991} 
\begin{equation}\label{eq:Qdef}
 Q_{\rm ss}(\alpha)=\frac{1}{\pi}\braket{\alpha|\tilde{\rho}_{\rm ss,\, c}|\alpha}=\frac{1}{\pi}e^{-|\alpha|^2}\sum_{n,m}\frac{\alpha^{*n}\alpha^m}{\sqrt{n!m!}} \braket{n|\rho_{\rm ss,\, c}|m},
\end{equation}
where
\begin{equation}
 \braket{n|\tilde{\rho}_{\rm ss,\, c}|m} \equiv \lim_{t \to \infty} {}_{a}\langle n|[{}_{A}\langle 2| \tilde{\rho}(t) |2\rangle _{A} + {}_{A}\langle 1| \tilde{\rho}(t) |1\rangle _{A}]|m\rangle_{a}
\end{equation}
and $\tilde{\rho}(t)$ is the system density operator in a frame rotating at $\omega_0$. Motivated by Figs. 2-4 of \cite{Alsing1991}, all of which plotted for a constant $g/\kappa=10$, we instead coherently drive the cavity mode such that a constant empty-cavity steady-state excitation amplitude $|\overline{\mathcal{E}}_0|/\kappa=10$ is maintained throughout while the system size parameter $n_{\rm sc}=\gamma^2/(8g^2)$ (the saturation photon number of absorptive optical bistability) changes from $0$ to $4.1 \times 10^{-3}$ in Panel I, and from $0$ to $5 \times 10^{-3}$ in Panel II of Fig. \ref{fig:Qfuncs}. For all the operating conditions used in this figure, the Maxwell-Bloch equations predict a single output state (according to the state equation for absorptive optical bistability \cite{Savage1988}) with $|\alpha_{\rm ss}|=|\overline{\mathcal{E}}_0|/\kappa=10$ \textemdash{the} steady-state cavity amplitude in the absence of the two-state atom \textemdash{while} the neoclassical equations predict $|\alpha_{\rm ss}|=0$, since $|\overline{\mathcal{E}}_0| \leq g/2$ in both panels. We also note that for $|\overline{\mathcal{E}}_0|/g \geq 1/(2\sqrt{2})$, which is the case for the drive amplitude used in Fig. \ref{fig:Qfuncs}, the Maxwell-Bloch equations in the limit $\gamma \to 0$ yield as well a steady state with the empty-cavity excitation (see Sec. 16.3.1 of \cite{QO2}).   

As we can observe in these distributions, the quantum picture differs substantially from the mean-field predictions: the steady-state intracavity excitation increases with $n_{\rm sc}$ while the field distribution always maintains a symmetry with respect to the real axis. The distribution peaks occur at complex conjugate amplitudes which are always bounded by $|\overline{\mathcal{E}}_0|/\kappa$ along the two orthogonal directions set by $\overline{\mathcal{E}}_0$; those peaks mark the attractors organizing symmetry breaking in the background of appreciable spontaneous emission. The excitation path joining the two symmetric states is eventually occupied with the largest excitation probability in a from of a single elongated peak inscribed along the ``bridge'' formed between the two branches of the JC split ladder, as we can see in frames (d) of both panels; the two branches are mixed by increasingly more frequent spontaneous emission events [see also Fig. 6 of \cite{Carmichael2015}, depicting complex-amplitude bimodality, as well as Fig. 3 of \cite{Kilin91} for one of the earliest accounts on phase bimodality]. While the excitation landscape have been set up below the critical point, as we see in Panel I, the occurrence of spontaneous dressed-state polarization at $|\overline{\mathcal{E}}_0|=g/2$ decreases substantially the vacuum-state occupation probability for every value of $n_{\rm sc}$. 

To get a better appreciation of the dynamics associated with spontaneous symmetry breaking, in Fig. \ref{fig:Qfuncevol} we plot the intracavity field distribution along the transient evolution leading to the steady state, when driving above the critical point and starting from the ground state of the JC oscillator. The distribution peaks follow a path constrained to the interior of the neoclassical curve parametrized by $(x(\overline{\varepsilon}_0), y(\overline{\varepsilon}_0))$ with $x(\overline{\varepsilon}_0)=(\overline{\varepsilon}_0/\kappa) \{1-[g/(2\overline{\varepsilon}_0)]^2\}$ and $y(\overline{\varepsilon}_0)=\pm [g/(2\kappa)] \sqrt{1-[g/(2\overline{\varepsilon}_0)]^2}$, where $0\leq \overline{\varepsilon}_0 \leq |\overline{\mathcal{E}}_0|$ (here we take $\overline{\mathcal{E}}_0=i|\overline{\mathcal{E}}_0|$). In this transient manifestation of quantum criticality, spontaneous symmetry breaking occurs from the very beginning of the evolution to the steady-state distribution depicted in the inset. In fact, the later stages of this evolution which are not imprinted on the phase space primarily establish the role of spontaneous emission in promoting switching between the two ladders; the two main peaks occur almost at the same position but are significantly elongated along the excitation pathway. 

\section{Concluding remarks}

In conclusion, we have investigated various aspects of the coherence associated with strong resonant light-matter interaction when transitioning from the weak excitation regime to the onset of a dissipative quantum phase transition of second order. For weak excitation, $2(|\overline{\mathcal{E}}_0|/g) \ll 1$, all calculated observables and correlation functions are directly expressed in terms of the degree of squeezing of the intracavity field. The incoherent spectra for both forwards and side-scattered light are sums of the same squared Lorentzian distributions \textemdash{an} evidence of fluctuation squeezing. Moreover, the decoherence effects due to cavity photon loss and spontaneous emission are placed on equal footing, since the correlation functions of first and second order are only functions of the sum $\kappa + \gamma/2$. This is the case for both the (incoherent) fluorescence spectrum the spectrum of the transmitted light. 

For stronger driving, however, when the higher ranks of the two branches forming the split JC ladder are being populated, the situation is markedly different. The two decoherence channels are assessed in terms of their ability to precipitate ladder-switching events, and spontaneous emission is special in that respect, as reflected in the role of the system-size parameter \textemdash{a} function of the ratio $\gamma/g$. Pictorially, what we are in essence dealing with is the behavior of quantum fluctuations along the line of zero detuning ($\Delta \omega=0$) in visualizations of out-of-equilibrium quantum dynamics like Fig. 1 or Fig. 2(a) of \cite{Carmichael2015}. Those visualizations are shaped in a fundamental way by the inputs and outputs in the scattering configuration modeled by the JC oscillator. It remains for the experiment to tie the various pieces together while tracing this line, from the squeezing-induced linewidth narrowing of the weak-excitation regime to the occurrence of spontaneous symmetry breaking at high excitation.

The data underlying this Communication are available at: \href{http://dx.doi.org/10.17632/s5srj363gp.1}{http://dx.doi.org/10.17632/s5srj363gp.1}.

\begin{acknowledgments}
I thank P. Rabl and M. Schuler for providing an initial efficient program solving a system of linear differential equations as the matrix form of a related Lindblad master equation. I am also grateful to H. J. Carmichael for valuable discussions. This work was supported by the Swedish Research Council (VR) alongside the Knut and Alice Wallenberg foundation (KAW).
\end{acknowledgments}

\bibliography{bibliography}

\begin{thebibliography}{21}%
\makeatletter
\providecommand \@ifxundefined [1]{%
 \@ifx{#1\undefined}
}%
\providecommand \@ifnum [1]{%
 \ifnum #1\expandafter \@firstoftwo
 \else \expandafter \@secondoftwo
 \fi
}%
\providecommand \@ifx [1]{%
 \ifx #1\expandafter \@firstoftwo
 \else \expandafter \@secondoftwo
 \fi
}%
\providecommand \natexlab [1]{#1}%
\providecommand \enquote  [1]{``#1''}%
\providecommand \bibnamefont  [1]{#1}%
\providecommand \bibfnamefont [1]{#1}%
\providecommand \citenamefont [1]{#1}%
\providecommand \href@noop [0]{\@secondoftwo}%
\providecommand \href [0]{\begingroup \@sanitize@url \@href}%
\providecommand \@href[1]{\@@startlink{#1}\@@href}%
\providecommand \@@href[1]{\endgroup#1\@@endlink}%
\providecommand \@sanitize@url [0]{\catcode `\\12\catcode `\$12\catcode
  `\&12\catcode `\#12\catcode `\^12\catcode `\_12\catcode `\%12\relax}%
\providecommand \@@startlink[1]{}%
\providecommand \@@endlink[0]{}%
\providecommand \url  [0]{\begingroup\@sanitize@url \@url }%
\providecommand \@url [1]{\endgroup\@href {#1}{\urlprefix }}%
\providecommand \urlprefix  [0]{URL }%
\providecommand \Eprint [0]{\href }%
\providecommand \doibase [0]{http://dx.doi.org/}%
\providecommand \selectlanguage [0]{\@gobble}%
\providecommand \bibinfo  [0]{\@secondoftwo}%
\providecommand \bibfield  [0]{\@secondoftwo}%
\providecommand \translation [1]{[#1]}%
\providecommand \BibitemOpen [0]{}%
\providecommand \bibitemStop [0]{}%
\providecommand \bibitemNoStop [0]{.\EOS\space}%
\providecommand \EOS [0]{\spacefactor3000\relax}%
\providecommand \BibitemShut  [1]{\csname bibitem#1\endcsname}%
\let\auto@bib@innerbib\@empty
%</preamble>
\bibitem [{\citenamefont {{Savage}}\ and\ \citenamefont
  {{Carmichael}}(1988)}]{Savage1988}%
  \BibitemOpen
  \bibfield  {author} {\bibinfo {author} {\bibfnamefont {C.~M.}\ \bibnamefont
  {{Savage}}}\ and\ \bibinfo {author} {\bibfnamefont {H.~J.}\ \bibnamefont
  {{Carmichael}}},\ }\bibfield  {title} {\enquote {\bibinfo {title} {Single
  atom optical bistability},}\ }\href {\doibase 10.1109/3.7075} {\bibfield
  {journal} {\bibinfo  {journal} {IEEE Journal of Quantum Electronics}\
  }\textbf {\bibinfo {volume} {24}},\ \bibinfo {pages} {1495--1498} (\bibinfo
  {year} {1988})}\BibitemShut {NoStop}%
\bibitem [{\citenamefont {{Jaynes}}\ and\ \citenamefont
  {{Cummings}}(1963)}]{JCpaper1963}%
  \BibitemOpen
  \bibfield  {author} {\bibinfo {author} {\bibfnamefont {E.~T.}\ \bibnamefont
  {{Jaynes}}}\ and\ \bibinfo {author} {\bibfnamefont {F.~W.}\ \bibnamefont
  {{Cummings}}},\ }\bibfield  {title} {\enquote {\bibinfo {title} {Comparison
  of quantum and semiclassical radiation theories with application to the beam
  maser},}\ }\href {\doibase 10.1109/PROC.1963.1664} {\bibfield  {journal}
  {\bibinfo  {journal} {Proceedings of the IEEE}\ }\textbf {\bibinfo {volume}
  {51}},\ \bibinfo {pages} {89--109} (\bibinfo {year} {1963})}\BibitemShut
  {NoStop}%
\bibitem [{\citenamefont {Bishop}\ \emph {et~al.}(2009)\citenamefont {Bishop},
  \citenamefont {Chow}, \citenamefont {Koch}, \citenamefont {Houck},
  \citenamefont {Devoret}, \citenamefont {Thuneberg}, \citenamefont {Girvin},\
  and\ \citenamefont {Schoelkopf}}]{Bishop2009}%
  \BibitemOpen
  \bibfield  {author} {\bibinfo {author} {\bibfnamefont {Lev~S.}\ \bibnamefont
  {Bishop}}, \bibinfo {author} {\bibfnamefont {J.~M.}\ \bibnamefont {Chow}},
  \bibinfo {author} {\bibfnamefont {Jens}\ \bibnamefont {Koch}}, \bibinfo
  {author} {\bibfnamefont {A.~A.}\ \bibnamefont {Houck}}, \bibinfo {author}
  {\bibfnamefont {M.~H.}\ \bibnamefont {Devoret}}, \bibinfo {author}
  {\bibfnamefont {E.}~\bibnamefont {Thuneberg}}, \bibinfo {author}
  {\bibfnamefont {S.~M.}\ \bibnamefont {Girvin}}, \ and\ \bibinfo {author}
  {\bibfnamefont {R.~J.}\ \bibnamefont {Schoelkopf}},\ }\bibfield  {title}
  {\enquote {\bibinfo {title} {Nonlinear response of the vacuum rabi
  resonance},}\ }\href {\doibase 10.1038/nphys1154} {\bibfield  {journal}
  {\bibinfo  {journal} {Nature Physics}\ }\textbf {\bibinfo {volume} {5}},\
  \bibinfo {pages} {105--109} (\bibinfo {year} {2009})}\BibitemShut {NoStop}%
\bibitem [{\citenamefont {Fink}\ \emph {et~al.}(2008)\citenamefont {Fink},
  \citenamefont {G{\"o}ppl}, \citenamefont {Baur}, \citenamefont {Bianchetti},
  \citenamefont {Leek}, \citenamefont {Blais},\ and\ \citenamefont
  {Wallraff}}]{Fink2008}%
  \BibitemOpen
  \bibfield  {author} {\bibinfo {author} {\bibfnamefont {J.~M.}\ \bibnamefont
  {Fink}}, \bibinfo {author} {\bibfnamefont {M.}~\bibnamefont {G{\"o}ppl}},
  \bibinfo {author} {\bibfnamefont {M.}~\bibnamefont {Baur}}, \bibinfo {author}
  {\bibfnamefont {R.}~\bibnamefont {Bianchetti}}, \bibinfo {author}
  {\bibfnamefont {P.~J.}\ \bibnamefont {Leek}}, \bibinfo {author}
  {\bibfnamefont {A.}~\bibnamefont {Blais}}, \ and\ \bibinfo {author}
  {\bibfnamefont {A.}~\bibnamefont {Wallraff}},\ }\bibfield  {title} {\enquote
  {\bibinfo {title} {Climbing the jaynes--cummings ladder and observing its
  nonlinearity in a cavity qed system},}\ }\href {\doibase 10.1038/nature07112}
  {\bibfield  {journal} {\bibinfo  {journal} {Nature}\ }\textbf {\bibinfo
  {volume} {454}},\ \bibinfo {pages} {315--318} (\bibinfo {year}
  {2008})}\BibitemShut {NoStop}%
\bibitem [{\citenamefont {Hamsen}\ \emph {et~al.}(2017)\citenamefont {Hamsen},
  \citenamefont {Tolazzi}, \citenamefont {Wilk},\ and\ \citenamefont
  {Rempe}}]{Hamsen2017}%
  \BibitemOpen
  \bibfield  {author} {\bibinfo {author} {\bibfnamefont {Christoph}\
  \bibnamefont {Hamsen}}, \bibinfo {author} {\bibfnamefont {Karl~Nicolas}\
  \bibnamefont {Tolazzi}}, \bibinfo {author} {\bibfnamefont {Tatjana}\
  \bibnamefont {Wilk}}, \ and\ \bibinfo {author} {\bibfnamefont {Gerhard}\
  \bibnamefont {Rempe}},\ }\bibfield  {title} {\enquote {\bibinfo {title}
  {Two-photon blockade in an atom-driven cavity qed system},}\ }\href {\doibase
  10.1103/PhysRevLett.118.133604} {\bibfield  {journal} {\bibinfo  {journal}
  {Phys. Rev. Lett.}\ }\textbf {\bibinfo {volume} {118}},\ \bibinfo {pages}
  {133604} (\bibinfo {year} {2017})}\BibitemShut {NoStop}%
\bibitem [{\citenamefont {Carmichael}(2015)}]{Carmichael2015}%
  \BibitemOpen
  \bibfield  {author} {\bibinfo {author} {\bibfnamefont {H.~J.}\ \bibnamefont
  {Carmichael}},\ }\bibfield  {title} {\enquote {\bibinfo {title} {Breakdown of
  photon blockade: A dissipative quantum phase transition in zero
  dimensions},}\ }\href {\doibase 10.1103/PhysRevX.5.031028} {\bibfield
  {journal} {\bibinfo  {journal} {Phys. Rev. X}\ }\textbf {\bibinfo {volume}
  {5}},\ \bibinfo {pages} {031028} (\bibinfo {year} {2015})}\BibitemShut
  {NoStop}%
\bibitem [{\citenamefont {Alsing}\ and\ \citenamefont
  {Carmichael}(1991)}]{Alsing1991}%
  \BibitemOpen
  \bibfield  {author} {\bibinfo {author} {\bibfnamefont {P}~\bibnamefont
  {Alsing}}\ and\ \bibinfo {author} {\bibfnamefont {H~J}\ \bibnamefont
  {Carmichael}},\ }\bibfield  {title} {\enquote {\bibinfo {title} {Spontaneous
  dressed-state polarization of a coupled atom and cavity mode},}\ }\href
  {\doibase 10.1088/0954-8998/3/1/003} {\bibfield  {journal} {\bibinfo
  {journal} {Quantum Opt.}\ }\textbf {\bibinfo {volume} {3}},\ \bibinfo {pages}
  {13--32} (\bibinfo {year} {1991})}\BibitemShut {NoStop}%
\bibitem [{\citenamefont {Armen}\ \emph {et~al.}(2009)\citenamefont {Armen},
  \citenamefont {Miller},\ and\ \citenamefont {Mabuchi}}]{Armen2009}%
  \BibitemOpen
  \bibfield  {author} {\bibinfo {author} {\bibfnamefont {Michael~A.}\
  \bibnamefont {Armen}}, \bibinfo {author} {\bibfnamefont {Anthony~E.}\
  \bibnamefont {Miller}}, \ and\ \bibinfo {author} {\bibfnamefont {Hideo}\
  \bibnamefont {Mabuchi}},\ }\bibfield  {title} {\enquote {\bibinfo {title}
  {Spontaneous dressed-state polarization in the strong driving regime of
  cavity qed},}\ }\href {\doibase 10.1103/PhysRevLett.103.173601} {\bibfield
  {journal} {\bibinfo  {journal} {Phys. Rev. Lett.}\ }\textbf {\bibinfo
  {volume} {103}},\ \bibinfo {pages} {173601} (\bibinfo {year}
  {2009})}\BibitemShut {NoStop}%
\bibitem [{\citenamefont {Mabuchi}\ and\ \citenamefont
  {Wiseman}(1998)}]{Mabuchi1998}%
  \BibitemOpen
  \bibfield  {author} {\bibinfo {author} {\bibfnamefont {H.}~\bibnamefont
  {Mabuchi}}\ and\ \bibinfo {author} {\bibfnamefont {H.~M.}\ \bibnamefont
  {Wiseman}},\ }\bibfield  {title} {\enquote {\bibinfo {title} {Retroactive
  quantum jumps in a strongly coupled atom-field system},}\ }\href {\doibase
  10.1103/PhysRevLett.81.4620} {\bibfield  {journal} {\bibinfo  {journal}
  {Phys. Rev. Lett.}\ }\textbf {\bibinfo {volume} {81}},\ \bibinfo {pages}
  {4620--4623} (\bibinfo {year} {1998})}\BibitemShut {NoStop}%
\bibitem [{\citenamefont {Carmichael}(2008)}]{QO2}%
  \BibitemOpen
  \bibfield  {author} {\bibinfo {author} {\bibfnamefont {H.~J.}\ \bibnamefont
  {Carmichael}},\ }\href@noop {} {\emph {\bibinfo {title} {Statistical Methods
  in Quantum Optics 2: Non-Classical Fields}}}\ (\bibinfo  {publisher}
  {Springer, Berlin},\ \bibinfo {year} {2008})\ Chap.~\bibinfo {chapter}
  {16}\BibitemShut {NoStop}%
\bibitem [{\citenamefont {Tan}(1999)}]{Tan1999}%
  \BibitemOpen
  \bibfield  {author} {\bibinfo {author} {\bibfnamefont {Sze~M}\ \bibnamefont
  {Tan}},\ }\bibfield  {title} {\enquote {\bibinfo {title} {A computational
  toolbox for quantum and atomic optics},}\ }\href {\doibase
  10.1088/1464-4266/1/4/312} {\bibfield  {journal} {\bibinfo  {journal}
  {Journal of Optics B: Quantum and Semiclassical Optics}\ }\textbf {\bibinfo
  {volume} {1}},\ \bibinfo {pages} {424--432} (\bibinfo {year}
  {1999})}\BibitemShut {NoStop}%
\bibitem [{\citenamefont {Alsing}\ \emph {et~al.}(1992)\citenamefont {Alsing},
  \citenamefont {Guo},\ and\ \citenamefont {Carmichael}}]{DynamicStarkEffect}%
  \BibitemOpen
  \bibfield  {author} {\bibinfo {author} {\bibfnamefont {P.}~\bibnamefont
  {Alsing}}, \bibinfo {author} {\bibfnamefont {D.-S.}\ \bibnamefont {Guo}}, \
  and\ \bibinfo {author} {\bibfnamefont {H.~J.}\ \bibnamefont {Carmichael}},\
  }\bibfield  {title} {\enquote {\bibinfo {title} {Dynamic stark effect for the
  jaynes-cummings system},}\ }\href {\doibase 10.1103/PhysRevA.45.5135}
  {\bibfield  {journal} {\bibinfo  {journal} {Phys. Rev. A}\ }\textbf {\bibinfo
  {volume} {45}},\ \bibinfo {pages} {5135--5143} (\bibinfo {year}
  {1992})}\BibitemShut {NoStop}%
\bibitem [{\citenamefont {Carmichael}\ \emph {et~al.}(1994)\citenamefont
  {Carmichael}, \citenamefont {Tian}, \citenamefont {Ren},\ and\ \citenamefont
  {Alsing}}]{cavityQEDBerman}%
  \BibitemOpen
  \bibfield  {author} {\bibinfo {author} {\bibfnamefont {H.~J.}\ \bibnamefont
  {Carmichael}}, \bibinfo {author} {\bibfnamefont {L.}~\bibnamefont {Tian}},
  \bibinfo {author} {\bibfnamefont {W.}~\bibnamefont {Ren}}, \ and\ \bibinfo
  {author} {\bibfnamefont {P.}~\bibnamefont {Alsing}},\ }\bibfield  {title}
  {\enquote {\bibinfo {title} {Nonperturbative atom-photon interactions in an
  optical cavity},}\ }in\ \href@noop {} {\emph {\bibinfo {booktitle} {Cavity
  Quantum Electrodynamics}}},\ \bibinfo {editor} {edited by\ \bibinfo {editor}
  {\bibfnamefont {P.~R.}\ \bibnamefont {Berman}}}\ (\bibinfo  {publisher}
  {Academic, Boston},\ \bibinfo {year} {1994})\ pp.\ \bibinfo {pages}
  {381--423}\BibitemShut {NoStop}%
\bibitem [{\citenamefont {Mandel}(1982)}]{Mandel1982}%
  \BibitemOpen
  \bibfield  {author} {\bibinfo {author} {\bibfnamefont {L.}~\bibnamefont
  {Mandel}},\ }\bibfield  {title} {\enquote {\bibinfo {title} {Squeezed states
  and sub-poissonian photon statistics},}\ }\href {\doibase
  10.1103/PhysRevLett.49.136} {\bibfield  {journal} {\bibinfo  {journal} {Phys.
  Rev. Lett.}\ }\textbf {\bibinfo {volume} {49}},\ \bibinfo {pages} {136--138}
  (\bibinfo {year} {1982})}\BibitemShut {NoStop}%
\bibitem [{\citenamefont {Carmichael}(1999)}]{QO1}%
  \BibitemOpen
  \bibfield  {author} {\bibinfo {author} {\bibfnamefont {H.~J.}\ \bibnamefont
  {Carmichael}},\ }\href@noop {} {\emph {\bibinfo {title} {Statistical Methods
  in Quantum Optics 1: Master Equations and Fokker-Planck Equations}}}\
  (\bibinfo  {publisher} {Springer, Berlin},\ \bibinfo {year} {1999})\
  Chap.~\bibinfo {chapter} {2}\BibitemShut {NoStop}%
\bibitem [{\citenamefont {Shamailov}\ \emph {et~al.}(2010)\citenamefont
  {Shamailov}, \citenamefont {Parkins}, \citenamefont {Collett},\ and\
  \citenamefont {Carmichael}}]{Shamailov2010}%
  \BibitemOpen
  \bibfield  {author} {\bibinfo {author} {\bibfnamefont {S.S.}\ \bibnamefont
  {Shamailov}}, \bibinfo {author} {\bibfnamefont {A.S.}\ \bibnamefont
  {Parkins}}, \bibinfo {author} {\bibfnamefont {M.J.}\ \bibnamefont {Collett}},
  \ and\ \bibinfo {author} {\bibfnamefont {H.J.}\ \bibnamefont {Carmichael}},\
  }\bibfield  {title} {\enquote {\bibinfo {title} {Multi-photon blockade and
  dressing of the dressed states},}\ }\href {\doibase
  https://doi.org/10.1016/j.optcom.2009.10.062} {\bibfield  {journal} {\bibinfo
   {journal} {Optics Communications}\ }\textbf {\bibinfo {volume} {283}},\
  \bibinfo {pages} {766--772} (\bibinfo {year} {2010})},\ \bibinfo {note}
  {{Q}uo vadis Quantum Optics?}\BibitemShut {Stop}%
\bibitem [{\citenamefont {Carmichael}\ \emph {et~al.}(1991)\citenamefont
  {Carmichael}, \citenamefont {Brecha},\ and\ \citenamefont
  {Rice}}]{Carmichael1991}%
  \BibitemOpen
  \bibfield  {author} {\bibinfo {author} {\bibfnamefont {H.J.}\ \bibnamefont
  {Carmichael}}, \bibinfo {author} {\bibfnamefont {R.J.}\ \bibnamefont
  {Brecha}}, \ and\ \bibinfo {author} {\bibfnamefont {P.R.}\ \bibnamefont
  {Rice}},\ }\bibfield  {title} {\enquote {\bibinfo {title} {Quantum
  interference and collapse of the wavefunction in cavity qed},}\ }\href
  {\doibase https://doi.org/10.1016/0030-4018(91)90194-I} {\bibfield  {journal}
  {\bibinfo  {journal} {Optics Communications}\ }\textbf {\bibinfo {volume}
  {82}},\ \bibinfo {pages} {73--79} (\bibinfo {year} {1991})}\BibitemShut
  {NoStop}%
\bibitem [{\citenamefont {Chough}\ and\ \citenamefont
  {Carmichael}(1996)}]{Chough1996}%
  \BibitemOpen
  \bibfield  {author} {\bibinfo {author} {\bibfnamefont {Y.~T.}\ \bibnamefont
  {Chough}}\ and\ \bibinfo {author} {\bibfnamefont {H.~J.}\ \bibnamefont
  {Carmichael}},\ }\bibfield  {title} {\enquote {\bibinfo {title} {Nonlinear
  oscillator behavior in the jaynes-cummings model},}\ }\href {\doibase
  10.1103/PhysRevA.54.1709} {\bibfield  {journal} {\bibinfo  {journal} {Phys.
  Rev. A}\ }\textbf {\bibinfo {volume} {54}},\ \bibinfo {pages} {1709--1714}
  (\bibinfo {year} {1996})}\BibitemShut {NoStop}%
\bibitem [{\citenamefont {{Rice}}\ and\ \citenamefont
  {{Carmichael}}(1988)}]{RiceCarmichaelIEEE}%
  \BibitemOpen
  \bibfield  {author} {\bibinfo {author} {\bibfnamefont {P.~R.}\ \bibnamefont
  {{Rice}}}\ and\ \bibinfo {author} {\bibfnamefont {H.~J.}\ \bibnamefont
  {{Carmichael}}},\ }\bibfield  {title} {\enquote {\bibinfo {title}
  {Single-atom cavity-enhanced absorption. i. photon statistics in the
  bad-cavity limit},}\ }\href {\doibase 10.1109/3.974} {\bibfield  {journal}
  {\bibinfo  {journal} {IEEE Journal of Quantum Electronics}\ }\textbf
  {\bibinfo {volume} {24}},\ \bibinfo {pages} {1351--1366} (\bibinfo {year}
  {1988})}\BibitemShut {NoStop}%
\bibitem [{\citenamefont {Carmichael}(1993)}]{Carmichael1993}%
  \BibitemOpen
  \bibfield  {author} {\bibinfo {author} {\bibfnamefont {H.~J.}\ \bibnamefont
  {Carmichael}},\ }\bibfield  {title} {\enquote {\bibinfo {title} {Quantum
  trajectory theory for cascaded open systems},}\ }\href {\doibase
  10.1103/PhysRevLett.70.2273} {\bibfield  {journal} {\bibinfo  {journal}
  {Phys. Rev. Lett.}\ }\textbf {\bibinfo {volume} {70}},\ \bibinfo {pages}
  {2273--2276} (\bibinfo {year} {1993})}\BibitemShut {NoStop}%
\bibitem [{\citenamefont {Kilin}\ and\ \citenamefont
  {Krinitskaya}(1991)}]{Kilin91}%
  \BibitemOpen
  \bibfield  {author} {\bibinfo {author} {\bibfnamefont {S.~Ya.}\ \bibnamefont
  {Kilin}}\ and\ \bibinfo {author} {\bibfnamefont {T.~B.}\ \bibnamefont
  {Krinitskaya}},\ }\bibfield  {title} {\enquote {\bibinfo {title} {Single-atom
  phase bistability in a fundamental model of quantum optics},}\ }\href
  {\doibase 10.1364/JOSAB.8.002289} {\bibfield  {journal} {\bibinfo  {journal}
  {J. Opt. Soc. Am. B}\ }\textbf {\bibinfo {volume} {8}},\ \bibinfo {pages}
  {2289--2295} (\bibinfo {year} {1991})}\BibitemShut {NoStop}%
\end{thebibliography}%

\end{document}